\documentclass[3p,amsmath,amssymb,compress,twocolumn]{elsarticle}
\usepackage{amsmath}
\usepackage{graphicx}
\usepackage{latexsym}
\usepackage{amsmath,amssymb}
\usepackage{bm} 
\usepackage{xcolor}
\usepackage{stackrel}
\usepackage{accents}
\usepackage{braket}

\journal{Annals of Physics: contribution to the Igor E. Dzyaloshinskii Memorial Special Issue}

\numberwithin{equation}{section}

\begin{document}

\begin{frontmatter}	

\title{Direct current in a stirred optical lattice}

\author[UNH]{Sergey S. Pershoguba}
\author[UMD]{Victor M. Yakovenko}
\address[UNH]{Department of Physics and Astronomy, University of New Hampshire, Durham, New Hampshire 03824, USA}
\address[UMD]{JQI, Department of Physics, University of Maryland, College Park, Maryland 20742, USA}

\begin{abstract}
We study how the energy dispersion of bosonic atoms loaded into an optical lattice becomes modified due to periodic circular stirring of the lattice to the second order in the strength of stirring.  If the lattice breaks mirror symmetry, the bosonic atoms may acquire a nonzero group velocity at the center of the Brillouin zone and produce a nonzero direct current.  This effect is similar to the circular photogalvanic effect in solid-state physics.  It can be used to transport neutral bosonic atoms in an optical lattice over a given distance in an arbitrary direction.  However, when the drive frequency is detuned to avoid resonant transitions with energy absorption, we argue that the induced current is not persistent, but transient.  An experimental study of the induced current relaxation could give answers to perplexing questions about equilibrization in driven systems.
\end{abstract}

\end{frontmatter}


\section{Introduction} \label{sec:intro}
An applied electric field $\bm E(t)$, which is generally time-dependent, causes a shift of quantum energy levels. For localized systems with discrete levels, such as atoms or molecules \cite{Kobe1983,Delone1999,haas2006}, this effect is known as the Stark-Bloch-Siegert (SBS) energy shift.  But in extended systems, such as electrons in solids \cite{Sie2015,Sie2017}, there are additional contributions to the energy shift besides the conventional SBS shift.  This is because the wavefunctions in solids are delocalized and characterized by quasimomentum $\hbar\bm k$, which is not conserved in the presence of $\bm E(t)$.  A systematic derivation of the energy band renormalization in solids to the second order in $\bm E(t)$ was recently done in Ref.~\cite{PershogubaYakovenko2022}.  The paper identified a term that couples the Berry curvature $\bm\Omega(\bm k)$ of the electrons in a crystal and the helicity $\bm h$ of circularly polarized light.  Based on this coupling, Ref.~\cite{PershogubaYakovenko2022} proposed optical control of the sign of spontaneous orbital magnetization in the recently discovered Chern insulators in Moir\'e materials \cite{Andrei2021}.

Here we discuss further physical implications of the second-order energy shift $\varepsilon^{(2)}(\bm k)$ derived in Ref.~\cite{PershogubaYakovenko2022} and evaluate the corresponding correction to the group velocity of quasiparticles.  Under suitable conditions, it may result in a direct current (DC) induced by circularly rotating electric field $\bm E(t)$.  This is known as the circular photogalvanic effect in solid-state physics and has been extensively studied in the literature for semiconductors and insulators \cite{FridkinBook,BelinicherReview,SturmanFridkinBook}.  This effect was mostly discussed, e.g.,\ in Refs.~\cite{Belinicher1986,Sipe2000,Rappe2012,RostamiPolini2018,Moore2017,Moore2019,Fregoso2019,Barik2020,Grushin2020,Yanase2021}, for resonant transitions between energy bands, when the driving frequency $\omega$ is greater than the energy gap.  In this case, a steady DC current is possible, e.g.,\ due to asymmetric population of the upper and lower bands \cite{Belinicher1986}.

In contrast, here we focus on the opposite case where $\omega$ is smaller than the gap, so resonant transitions with photon absorption are not possible, but quantum states in different bands form coherent superpositions due to the presence of $\bm E(t)$.  For a fully occupied band in an electronic insulator, the total DC current due to any modification of group velocity vanishes, because it is given by an integral in momentum space of a full derivative.  (However, Ref.~\cite{Kaplan2020} arrived to a different conclusion due to lifetime effects.)  For a partially occupied band, the situation is more subtle.  Some papers claim that the induced DC current is nonzero in this case \cite{MooreOrenstein2010,SodemannFu2015,Nagaosa2022,Sodemann2022}.  In contrast, we argue that the current vanishes in thermal equilibrium, but can be nonzero during a transition to equilibrium, in agreement with Ref.~\cite{Belinicher1986}.  The difference in conclusions originates from the ambiguity in assumptions about thermal equilibrium for a driven system.  We focus on the transient current appearing when the circular drive is turned on and kept steady, whereas Ref.~\cite{Belinicher1986} considered periodic modulation in time of the intensity or degree of polarization of the driving field.

In contrast to previous literature, we study an analog of the circular photogalvanic effect for neutral bosonic atoms in an optical lattice, where they occupy the lowest energy state with $\bm k=0$ at low temperature, unlike fermions in electronic materials.  (We assume that temperature is above the Bose-Einstein condensation and use a simple quasiparticle description without superfluid condensate.)  An analogue of the electric force $e\bm E(t)$ for neutral atoms was experimentally realized in Ref.~\cite{Jotzu2014} by shaking of the optical lattice, which produces the force of inertia
\begin{equation}  \label{force}
  \bm F(t)=-M\,\frac{d^2\bm R(t)}{dt^2}.
\end{equation}
Here $M$ is the mass of an atom, and the time-dependent vector $\bm R(t)$ represents displacement of the optical lattice.  We show that circular shaking (stirring) of the lattice may shift the energy minimum away from $\bm k=0$, so the atoms with $\bm k=0$ would acquire a nonzero group velocity $\bm V$ and thus carry current.  This current is not persistent, but transient and eventually vanishes when the atoms relax to a new energy minimum at $\bm k\neq0$ due to dissipation.  Nevertheless, if relaxation time is sufficiently long, the induced current may be used for transporting neutral bosonic atoms in an optical lattice over a fixed distance by stirring for a fixed time.  Moreover, measuring relaxation of the transient current may provide information about dissipation processes in cold bosonic gases.

In contrast to electronic materials, resonant transitions with photon absorption are less relevant for bosons in an optical lattice.  First, resonant absorption requires tuning of the driving frequency $\omega$ to the energy band gap at $\bm k=0$, whereas a nonresonant energy shift is obtained for a generic frequency.  Second, the unrenormalized group velocity vanishes at $\bm k=0$ by time-reversal symmetry, so a change in population due to interband transitions would not produce current.  The current is induced only due to renormalization of the group velocity $\bm V$ proportional to the gradient $\partial\bm\Omega(\bm k)/\partial\bm k$ of the Berry curvature in momentum space, which may be nonzero at $\bm k=0$, even though $\bm\Omega(\bm k=0)=0$.

\begin{figure}
\begin{center}
\includegraphics[width=0.47\textwidth]{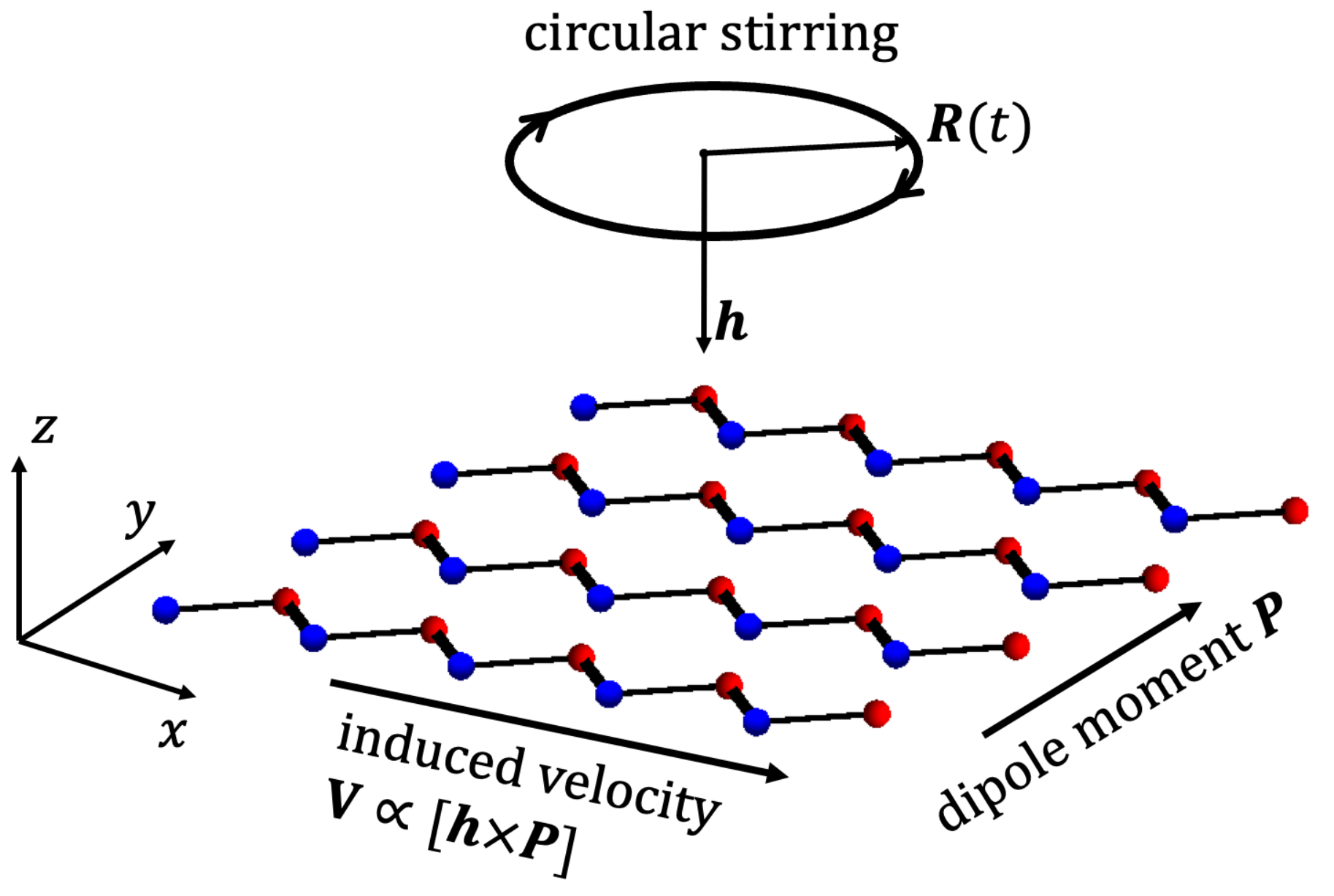} 
\caption{A 2D optical lattice consisting of 1D zigzag chains with alternating on-site energies.  The induced velocity $\bm V$ is along the chains and perpendicular to the helicity vector $\bm h$ of circular stirring and to the dipole moment $\bm P$ of the lattice.}
\label{fig:geometry}
\end{center}
\end{figure}

We study the geometry where bosonic atoms are loaded into a two-dimensional (2D) optical lattice in the $xy$ plane, as shown in Fig.~\ref{fig:geometry}.  Stirring of the lattice is described by the in-plane displacement vector $\bm R(t)$.  The direction of stirring is characterized by the out-of-plane helicity vector ${\bm h \parallel \hat{\bm z}}$, which is defined by Eq.~(\ref{helicity}) and points up or down for counterclockwise or clockwise stirring.  A nonzero current can be obtained when the optical lattice breaks mirror symmetry and admits a polar vector $\bm P$.  In the case of decoupled one-dimensional (1D) zigzag chains with alternating on-site energies, shown in red and blue in Fig.~\ref{fig:geometry}, the vector $\bm P$ is the in-plane dipole moment perpendicular to the chains.  Then the induced velocity $\bm V$ is along the chains and perpendicular to both $\bm h$ and $\bm P$:
\begin{align}
    \bm V \propto [\bm h \times \bm P]. \label{hXP}
\end{align}
Since $\bm h$ is an axial time-reversal-odd vector similar to magnetic field, while $\bm P$ is a polar and time-reversal-even vector similar to electric field, their vector product has the symmetry of a velocity $\bm V$ similar to the Hall current.  The 1D geometry shown in Fig.~\ref{fig:geometry} is a limiting case of the 2D honeycomb lattice with anisotropic couplings between the sites studied experimentally in Ref.~\cite{Jotzu2014}.

The paper is structured as follows. In Sec.~\ref{sec:non_inertial}, we derive the effect of shaking in the non-inertial reference frame of the lattice.  In Sec.~\ref{sec:review}, we discuss  renormalization of the energy dispersion to the second order in $\bm F$ following Ref.~\cite{PershogubaYakovenko2022}. After introducing an anisotropic honeycomb lattice in Sec.~\ref{sec:tb_model}, we evaluate the velocity $\bm V$ in the 1D and 2D limits of the model in Secs.~\ref{sec:1d_limit} and \ref{sec:2D_limit}, respectively. Conclusions are given in Sec.~\ref{sec:conclusions}.  Technical details of the derivations are given in  \ref{sec:berry_curv_appendix} and \ref{sec:perturbative}.  In \ref{sec:Hellmann-Feynman}, we prove a generalization of the Hellmann-Feynman theorem for the Floquet states.  It demonstrates that the time-averaged expectation value of the velocity operator is equal to the group velocity derived from the renormalized energy dispersion.  In \ref{sec:kinetic}, we present a direct calculation of the induced DC current using the quantum kinetic equation popular in the literature \cite{Sipe2000}.  It gives the same answer as the derivation based on the renormalized group velocity.

\section{Shaking of the optical lattice} \label{sec:non_inertial}
We start with analyzing the effect of periodic shaking of an optical lattice.  The wavefunction $\psi_{\rm lab}(\bm r,t)$ of an atom of mass $M$ in the laboratory reference frame is governed by the time-dependent Schr\"odinger equation
\begin{align}
    i\hbar\,\partial_t \psi_{\rm lab}(\bm r,t) 
    = \left\{\frac{\hat{\bm p}^2}{2M} + U[\bm r-\bm R(t)]\right\} \psi_{\rm lab}(\bm r,t). 
    \label{schrod}
\end{align}
Here $\hat{\bm p} = -i\hbar\,\partial/\partial \bm r$ is the momentum operator, and $U[\bm r-\bm R(t)]$ represents the lattice potential shifted by the displacement vector $\bm R(t)$.  The wavefunction $\psi_{\rm lat}(\bm r,t)$ in the noninertial reference frame of the moving lattice is related to the wavefunction in the inertial laboratory frame by the displacement operator and a gauge transformation \cite{LandauVol3}:
\begin{align}
  &\psi_{\rm lab}(\bm r,t) = e^{-i\hat{\bm p}\cdot\bm R(t)/\hbar}\, 
  e^{i\,\varphi(\bm r,t)} \, \psi_{\rm lat}(\bm r,t), 
  \label{Galileo}\\
&\hbar\,\varphi(\bm r,t) = M \bm r\cdot\dot{\bm R}(t)  + \frac M2 \int^t dt' \dot {\bm R}^2(t').
\end{align}
Substituting Eq.~(\ref{Galileo}) into Eq.~(\ref{schrod}), we obtain the time-dependent Schr\"odinger equation in the lattice reference frame
\begin{align}
    &i\hbar\,\partial_t \psi_{\rm lat}(\bm r,t) 
    = \left[\frac{\bm p^2}{2M}  + U(\bm r) - \bm F(t)\cdot \bm r \right] 
    \psi_{\rm lat}(\bm r,t). \label{schrod_non_inert}
\end{align}
As expected, the effect of shaking enters via the time-dependent force of inertia $\bm F(t)$ given by Eq.~(\ref{force}). In the rest of the paper, we work with Eq.~(\ref{schrod_non_inert}) and drop the subscript in $\psi_{\rm lat}(\bm r,t)$ to simplify notation.

Equation~(\ref{schrod_non_inert}) is equivalent to that of a particle with charge $e$ in a time-dependent electric field $\bm E(t) =\bm F(t)/e$.  It was studied in Ref.~\cite{PershogubaYakovenko2022} for a monochromatic time dependence of the driving force at frequency $\omega$ 
\begin{align}
    \bm F(t) = \frac{\bm F(\omega)\, e^{-i\omega t}+ \bm F(- \omega)\, e^{i\omega t}}{2}, \label{harmonic_expansion}
\end{align} 
where the complex amplitudes are related as $\bm F(-\omega) = \bm F^\ast(\omega)$ to ensure that $\bm F(t)$ is real. The helicity $\bm h$ is defined as
\begin{align}
  \bm h ={\rm Im}\left[\bm F^\ast(\omega)\times\bm F(\omega)\right]
  \label{helicity}
\end{align}
and has the symmetry properties of magnetic field, because it is an axial time-reversal-odd vector.  In tensor form, Eq.~(\ref{helicity}) can be written as
\begin{align}
  {\rm Im}\left[F^\ast_\alpha(\omega)F_\beta(\omega)\right] 
  =\frac12 \epsilon_{\alpha\beta\gamma} h^\gamma,
  \label{antisymmetric}
\end{align}
where the spatial indices $\alpha$, $\beta$, and $\gamma$ take the values $x$, $y$, and $z$, whereas $\epsilon_{\alpha\beta\gamma}$ is the antisymmetric Levi-Civita tensor.

We focus on circular shaking (stirring) in the $xy$ plane, which is described by
\begin{align}
    \bm F(\omega) = (1,\pm i,0)\,F_{\pm}, 
    \label{F_amplitude}
\end{align}
where the signs $\pm$ correspond to clockwise or counterclockwise stirring.  The helicity vector $\bm h$ in this case points along $z$ axis
\begin{align}
  \bm h = \pm 2 |F_\pm|^2\, (0,0,1).
  \label{helicity+-}
\end{align}

\section{Renormalization of the energy spectrum due to time-periodic shaking}
\label{sec:review}
\subsection{The second-order correction} 

For the Schrodinger equation with periodic potential $U(\bm r)$,  the eigenstates are Bloch wavefunction labeled by the discrete band index $n$ and the continuous quasimomentum $\bm k$, which we simply call momentum for shortness. The wavefunctions for the energies $\varepsilon_n(\bm k)$ have the Bloch form $\psi_{n,\bm k} = e^{i\bm k\cdot\bm r} u_{n,\bm k}(\bm r)$, where $u_{n,\bm k}(\bm r)$ is periodic in $\bm r$.  

In the presence of the time-dependent force (\ref{schrod_non_inert}), it is practical~\cite{Bauer2020} to expand the wavefunction as
\begin{align} \label{psi-k(t)}
  \psi(t) = e^{i\bm r\cdot\tilde{\bm k}(t)} \sum_m c_{m,\bm k}(t) \, 
  e^{-\frac{i}{\hbar}\int\limits^t dt' \varepsilon_m[\tilde{\bm k}(t')]} \,
  u_{m,\tilde{\bm k}(t)}(\bm r),
\end{align} 
where the time-dependent momentum $\tilde{\bm k}(t)$ satisfies Newton's equation of motion
\begin{align} \label{Newton}
  \frac{d\tilde{\bm k}(t)}{dt}=\frac{\bm F(t)}{\hbar},
\end{align}
whereas $\bm k=\langle\tilde{\bm k}(t)\rangle_t$ is its time-averaged value.  Substituting Eq.~(\ref{psi-k(t)}) in the Schr\"odinger equation Eq.~(\ref{schrod_non_inert}), we obtain 
\begin{align} \label{c(t)-k(t)}
  i\hbar\dot c_{m,\bm k} = -\sum_{m'} F_\alpha(t)\, 
  r_{mm'}^\alpha(\tilde{\bm k})\,
  e^{\frac i\hbar\int\limits^t dt' \varepsilon_{mm'}(\tilde{\bm k})} c_{m',\bm k},
\end{align} 
where the matrix elements of the coordinate operator $\bm r$ in a crystal are expressed in terms of the Berry connections \cite{BlountBook1962,LandauBook}
\begin{align} \label{connections}
  r_{nm}^\alpha(\bm k) = i\langle u_{n,\bm k}| \frac{\partial}{\partial k_\alpha} | u_{m,\bm k}\rangle.
\end{align}
As in Ref.~\cite{PershogubaYakovenko2022}, we treat the monochromatic force (\ref{harmonic_expansion}) as a small perturbation in Eq.~(\ref{c(t)-k(t)}) and obtain the second-order correction $\varepsilon^{(2)}_n(\bm k)$ to eigenenergies 
\begin{equation}
\begin{aligned}  
  &  \varepsilon^{(2)}_n(\bm k) = \frac1{4\,(\hbar\omega)^2}\frac{\partial^2\varepsilon_n(\bm k)}{\partial k_\alpha \partial k_\beta }\,{\rm Re}\left[F^\ast_\alpha(\omega)F_\beta(\omega)\right]\, \\ & \,\, -\frac{\Omega_{n,\gamma}(\bm k)}{4\,\hbar\omega} \epsilon^{\alpha\beta\gamma} \,{\rm Im}\left[ F_\alpha^\ast(\omega)F_\beta(\omega)\right]  \\
  & \,\,- \frac{1}{4}\,{\rm Re}\sum_{m\neq n} \frac{r^{\alpha}_{nm}(\bm k)r^\beta_{mn}(\bm k)}{\varepsilon_{mn}(\bm k)-\hbar\omega}\,F_\alpha^\ast(\omega)F_\beta(\omega)   \\
  &\,\, - \frac{1}{4}\,{\rm Re}\sum_{m\neq n} \frac{\left[r^{\alpha}_{nm}(\bm k)r^\beta_{mn}(\bm k)\right]^\ast}{\varepsilon_{mn}(\bm k)+
  \hbar\omega} F_\alpha^\ast(\omega)F_\beta(\omega). \end{aligned}
  \label{full_stark_shift}
\end{equation}
Here we denote $\varepsilon_{mn}(\bm k) = \varepsilon_{m}(\bm k)-\varepsilon_{n}(\bm k)$ and introduce the Berry curvature vector
\begin{align}
 \bm \Omega_{n}(\bm k) = \frac{\partial}{\partial \bm k}  \times \bm r_{nn}(\bm k). 
 \label{berry_curv_def}
\end{align}
The subscripts of $\Omega_{n,\gamma}(\bm k)$ appearing in Eq.~(\ref{full_stark_shift}) label the band $n$ and the spatial component $\gamma$.  Let us briefly comment on the origin of the four terms in the right-hand side of Eq.~(\ref{full_stark_shift}). 

(i) 
The last two terms in Eq.~(\ref{full_stark_shift}) represent the conventional second-order perturbation theory with the interband matrix elements in the numerators and the energy denominators $\varepsilon_{mn}(\bm k)-\omega$ and $\varepsilon_{mn}(\bm k)+\omega$. These terms, originally derived for the systems with discrete energy spectrum, such as atoms and molecules \cite{Kobe1983,Delone1999,haas2006}, and known as the Stark and the Bloch-Siegert energy shifts, have been experimentally measured in various materials \cite{Sie2015,Sie2017}.

(ii) 
In contrast, the first two terms in Eq.~(\ref{full_stark_shift}) have intraband origin and occur only in extended systems, where the momentum $\tilde{\bm k}(t)$ is driven by the force $\bm F(t)$ according to Eq.~(\ref{Newton}). The first term in Eq.~(\ref{full_stark_shift}) comes from the renormalized energy $\langle\varepsilon_m[\tilde{\bm k}(t)]\rangle_t$ obtained by time averaging of the instantaneous energy $\varepsilon_m[\tilde{\bm k}(t)]$ in Eq.~(\ref{psi-k(t)}).  Representing $\tilde{\bm k}(t)$ as a sum of its time-averaged value $\bm k$ and a small deviation $\delta\bm k(t)$:
\begin{align}
    &\tilde {\bm k}(t) = \bm k + \delta \bm k(t),
    \label{delta-k}
\end{align}
we expand the instantaneous energy in terms of $\delta\bm k$ 
\begin{equation} \label{E(k)-expantion}
\begin{aligned}
\varepsilon_m[\tilde{\bm k}(t)] &\approx \varepsilon_m(\bm k)+\delta k_\alpha(t) \frac{\partial\varepsilon_m(\bm k)}{\partial k_\alpha} \\ 
 & \quad+\delta k_\alpha(t)\,\delta k_\beta(t) \,\frac12\frac{\partial^2\varepsilon_m(\bm k)}{\partial k_\alpha\partial k_\beta}.    \end{aligned}
\end{equation}
Using the monochromatic force (\ref{harmonic_expansion}) in Newton's equation (\ref{Newton}), we find
\begin{align}
    &\delta \bm k(t) = \frac{\bm F(- \omega)\, e^{i\omega t}-\bm F(\omega)\, e^{-i\omega t}}{2\,i\,\hbar\omega}.
    \label{k_vs_F}
\end{align}
Inserting Eq.~(\ref{k_vs_F}) into Eq.~(\ref{E(k)-expantion}) and averaging over time, we obtain the first term in Eq.~(\ref{full_stark_shift}).  This term was not written in Ref.~\cite{PershogubaYakovenko2022}, because it is symmetric with respect to the spatial indices $\alpha$ and $\beta$, whereas the paper focused on antisymmetric terms.

Importantly, this term, being proportional to the curvature tensor $\partial^2\varepsilon_n(\bm k)/\partial k_\alpha\partial k_\beta$, leads to flattening of the energy dispersion, because the energy at a given momentum $\bm k$ averages out with the energies at the neighboring momenta.  Therefore, applying a time-periodic force could be used as an experimental tool to tune flatness of the energy dispersion.  The latter is a crucial parameter controlling emergence of strongly-correlated phases, e.g.,\ in the recently-discovered Moir\'e materials \cite{Andrei2021}. 

Equation (\ref{E(k)-expantion}) also shows that the perturbation theory employed in our paper is applicable when the force $F$ is sufficiently small, so that the momentum deviation $\delta k \sim F/\hbar\omega \ll 1/a$ in Eq.~(\ref{k_vs_F}) is much smaller than the Brillouin zone size, where $a$ is the lattice spacing.  Thus, the limit $\omega\to0$ cannot be taken for a fixed $F$, and the nominal divergence at $\omega\to0$ in the first and the second terms of Eq.~(\ref{full_stark_shift}) is spurious.  Moreover, the square of $F$ in Eq.~(\ref{E(k)-expantion}) itself scales as $F^2\propto\omega^4$ according to Eq.~(\ref{force}).

(iii) 
The second term in Eq.~(\ref{full_stark_shift}) occurs due to accumulation of the Berry phase with time. For illustration, consider the rotating $\bm F$ described by Eq.~(\ref{F_amplitude}).  By Eq.~(\ref{k_vs_F}), it causes the momentum deviation $\delta\bm k(t)$ to move on a circle of radius $\delta k_R = F_+/\hbar\omega$. After a full cycle along the closed loop, the wavefunction accumulates the Berry phase 
\begin{align}
\phi &\approx \Omega_{n,z}(\bm k) \,\pi \,(\delta k_R)^2 \\
&= \Omega_{n,z}(\bm k) \,\pi \left( \frac{F_+}{\hbar\omega}\right)^2 \nonumber
\end{align}
proportional to the flux of the Berry curvature through the area enclosed by the circle.  Phase accumulation per period $T = 2\pi/\omega$ corresponds to the energy shift $\Delta\varepsilon_n =\hbar\,\phi/T$, which produces the second term in Eq.~(\ref{full_stark_shift}).  Given Eq.~(\ref{antisymmetric}), this term is proportional to the helicity $\bm h$ and vanishes for linear polarization.  To the best of our knowledge, this term was first identified in Ref.~\cite{PershogubaYakovenko2022}.

\subsection{Time-reversal symmetry and group velocity}
\label{sec:group-velocity}

The second-order correction (\ref{full_stark_shift}) modifies the energy dispersion and the group velocity
\begin{align}  
    &\tilde\varepsilon_n(\bm k) = \varepsilon_n(\bm k) + \varepsilon_n^{(2)}(\bm k),
    \label{renormalized_spectrum} \\
    &\tilde{\bm v}_n(\bm k)=\frac{\partial\tilde\varepsilon_n(\bm k)}{\partial\bm k}.
    \label{renormalized_velocity} 
\end{align}
The bilinear products ${\rm Re}\left[F^\ast_\alpha(\omega)F_\beta(\omega)\right]$ and ${\rm Im}\left[F^\ast_\alpha(\omega)F_\beta(\omega)\right]$ in Eq.~(\ref{full_stark_shift}) are symmetric and antisymmetric upon interchange of the spatial indices ${\alpha \leftrightarrow \beta}$ and upon time reversal.  Accordingly, the energy shift~(\ref{full_stark_shift}) can be split into the symmetric (s) and antisymmetric (a)  terms
\begin{align}
   & \varepsilon_n^{(2)}(\bm k) = \varepsilon_n^{(s)}(\bm k) + \varepsilon_n^{(a)}(\bm k). \label{two_band_full}
 \end{align}
Assuming that the bare Hamiltonian (without drive) has the time-reversal symmetry, the two terms in Eq.~(\ref{two_band_full}) are symmetric and antisymmetric with respect to time reversal (for a fixed drive) and the sign change of the momentum $\bm k$
\begin{align}
   & \varepsilon_n^{(s)}(-\bm k) = \varepsilon_n^{(s)}(\bm k), \quad 
   \varepsilon_n^{(a)}(-\bm k) = -\varepsilon_n^{(a)}(\bm k). 
   \label{k->-k}
 \end{align}

Generally, the current density $\bm j$ is given by an integral in momentum space of the occupation function $f_n(\bm k)$ and the group velocity $\tilde{\bm v}_n(\bm k)$
\begin{align}
  \bm j & = \sum_n\int \frac{d^2k}{(2\pi)^2}\,f_n(\bm k)\,\tilde{\bm v}_n(\bm k) 
  \label{current_density} \\
  & = \sum_n\int \frac{d^2k}{(2\pi)^2}\,f_n(\bm k)\,
  \frac{\partial\tilde\varepsilon_n(\bm k)}{\partial\bm k}.
  \label{current_group_velocity}
\end{align}
Integrating by parts over a periodic Brillouin zone, it can be identically rewritten as
\begin{align}
  \bm j = -\sum_n\int \frac{d^2k}{(2\pi)^2}\,\tilde\varepsilon_n(\bm k)\,
  \frac{\partial f_n(\bm k)}{\partial\bm k}.
  \label{current_energy}
\end{align}
Substituting Eq.~(\ref{full_stark_shift}) into Eq.~(\ref{current_energy}), we obtain the second-order contribution to the current, which is equivalent to Eq.~(7) in Ref.~\cite{Nagaosa2022} and similar equations in the papers cited therein.  Thus, the induced current derived in these papers is entirely due to modification of the group velocity in Eq.~(\ref{current_group_velocity}).  In particular, the nonlinear Hall effect discussed in Ref.~\cite{SodemannFu2015} originates from the second term in Eq.~(\ref{full_stark_shift}) that involves the Berry curvature.

Obviously, the current in Eq.~(\ref{current_energy}) vanishes for a fully occupied band, where $f_n(\bm k)=1$ for all $\bm k$, i.e.,\ for an insulator.  For a partially occupied band, the situation is more subtle.  If the occupation function $f[\tilde\varepsilon_n(\bm k)]$ depends only on the renormalized energy $\tilde\varepsilon_n(\bm k)$, the integral (\ref{current_group_velocity}) also vanishes identically.  Thus, when the system relaxes to thermal equilibrium at the modified energy spectrum $\tilde\varepsilon_n(\bm k)$, the steady current due to the modified group velocity $\tilde{\bm v}_n(\bm k)$ would vanish, as pointed out in Ref.~\cite{Belinicher1986}, in contrast to the conclusions of Refs.~\cite{SodemannFu2015,Nagaosa2022,Sodemann2022}.

Nevertheless, the current may be nonzero during the transient time after the drive is turned on, but before re-thermalization is achieved, while the occupation function $f[\varepsilon_n(\bm k)]$ still depends on the unmodified energy dispersion.  The unperturbed occupation function $f[\varepsilon_n(\bm k)]$ is an even function of $\bm k$, if the system without drive has the time-reversal symmetry.  Then, a nonzero transient current in Eq.~(\ref{current_group_velocity}) can only originate from the antisymmetric energy shift $\varepsilon_n^{(a)}(\bm k)$ in Eq.~(\ref{k->-k}), which is proportional to the helicity of the drive.

Technically, Eq.~(\ref{renormalized_spectrum}) gives a Floquet quasi-energy in the presence of a periodic-in-time drive.  It is not entirely obvious whether the system would eventually thermalize to the quasi-energy $\tilde\varepsilon_n(\bm k)$ in the presence of a steady drive, in contrast to the conventional energy $\varepsilon_n(\bm k)$ without drive.  Nevertheless, thermalization to $\tilde\varepsilon_n(\bm k)$ looks plausible and can be verified experimentally as discussed in Sec.~\ref{sec:1d_limit}.

Generally, when a system is driven at a frequency $\omega$, the induced current has several Fourier components. The linear response gives a current at the frequency $\omega$, whereas the second-order response produces a DC current at zero frequency and a second harmonic at $2\omega$.  The linear-response current does have an anomalous contribution besides the group velocity, which is ultimately responsible for the quantum Hall effect or an ac Hall effect at $\omega\neq0$, as discussed in many papers, e.g.,\ in Ref.~\cite{PershogubaYakovenko2022}.  One might have hoped for a similar anomalous contribution to the nonlinear response at a subgap frequency in the absence of resonant absorption.  However, we demonstrate in \ref{sec:Hellmann-Feynman} that it is not the case.  We prove a generalization of the Hellmann-Feynman theorem for the Floquet states and show that the nonlinear DC current in this regime is only given by the modified group velocity in Eq.~(\ref{current_group_velocity}).  To illustrate this statement, in \ref{sec:kinetic} we calculate the induced current directly, using the quantum kinetic equation for a two-band model, rather than implicitly by using the renormalized energy.  The two calculations give the same result.

\subsection{Application to bosons in an optical lattice} 

Given the conclusion that, for subgap illumination without resonant absorption, it is only possible to have transient current for a partially occupied band, we switch our focus to bosons in an optical lattice, as opposed to electrons in solids studied in previous literature.  At low temperature, the unperturbed occupation function for bosons $f_n(\bm k)\approx\delta^{(2)}(\bm k)$ can be approximated by a delta-function around $\bm k=0$ for the lowest energy band.  (While we assume that temperature is still above the Bose-Einstein condensation.)  Then the current in Eq.~(\ref{current_density}) is determined by the group velocity of the lowest band at $\bm k=0$, which we denote as $\bm V$:
\begin{align}
  &\bm V = \tilde{\bm v}_n(\bm k=0) 
  = \left.\frac{\partial\varepsilon_n^{(a)}(\bm k)}{\hbar\,\partial \bm k} \right|_{\bm k = 0}.
  \label{k=0}
\end{align}
A nonzero group velocity $\bm V$ for bosons in only possible when the driving force has nonzero helicity in Eq.~(\ref{antisymmetric}), thus breaking the time-reversal symmetry and enabling the antisymmetric energy shift $\varepsilon_n^{(a)}(\bm k)$.  In the next Section, we discuss this effect in more detail for a two-band model.

\section{A two-band model}

\subsection{A general two-band model}

In further presentation, we focus on a model with only two energy bands and label the lower band as $n$ and the upper as $m$.  In this case, the sums over $m$ in Eq.~(\ref{full_stark_shift}) contain only one term, and the antisymmetric part is related to the Berry curvature
\begin{align}
    {\rm Im}\,\left[r^\alpha_{nm}r^\beta_{mn}\right] = -\frac12 \epsilon^{\alpha\beta\gamma}\Omega_{n,\gamma}. 
    \label{Berry_relation}
\end{align}
Then the antisymmetric energy shift is proportional to the scalar product of the helicity $\bm h$ of the driving force and the Berry curvature $\bm\Omega_n(\bm k)$ of the lattice
\begin{align}  
  & \varepsilon_n^{(a)}(\bm k) = 
  -\frac{\varepsilon_{mn}^2(\bm k)}{\varepsilon^2_{mn}(\bm k)-(\hbar\omega)^2} \frac{\bm h\cdot\bm\Omega_n(\bm k)}{4\,\hbar\omega}.
  \label{two_band_asym}
\end{align}
Substituting Eq.~(\ref{two_band_asym}) into Eq.~(\ref{k=0}), we find that the group velocity $\bm V$ at $\bm k=0$ for the lower band $n$ is proportional to the gradient of the Berry curvature, while $\partial\varepsilon_{mn}(\bm k)/\partial\bm k=0$,
\begin{align}
  &\bm V =-\frac{\varepsilon_{mn}^2(0)}{\varepsilon^2_{mn}(0)-(\hbar\omega)^2}
  \frac{1}{4\,\hbar\omega}
  \left.\frac{\partial[\bm h\cdot\bm\Omega_n(\bm k)]}{\hbar\,\partial\bm k}\right|_{\bm k = 0}. 
  \label{vel}
\end{align}

The symmetric energy shift in Eq.~(\ref{full_stark_shift}) is
\begin{align}  
 & \varepsilon^{(s)}_n(\bm k) 
 = \,{\rm Re}\left[F^\ast_\alpha(\omega)F_\beta(\omega)\right] 
 \label{two_band_sym}
 \\ 
&\times\left\{\frac{1}{4\,(\hbar\omega)^2}\frac{\partial^2\varepsilon_n(\bm k)}{\partial k_\alpha\partial k_\beta}-\frac{\varepsilon_{mn}(\bm k)\,{\rm Re} \left[r^{\alpha}_{nm}(\bm k)r^\beta_{mn}(\bm k)\right]}{2\left[\varepsilon^2_{mn}(\bm k)-(\hbar\omega)^2\right]}\right\}.  \nonumber 
\end{align}

\subsection{The honeycomb lattice model} \label{sec:tb_model}

\begin{figure}
\begin{center}
\includegraphics[width=0.4\textwidth]{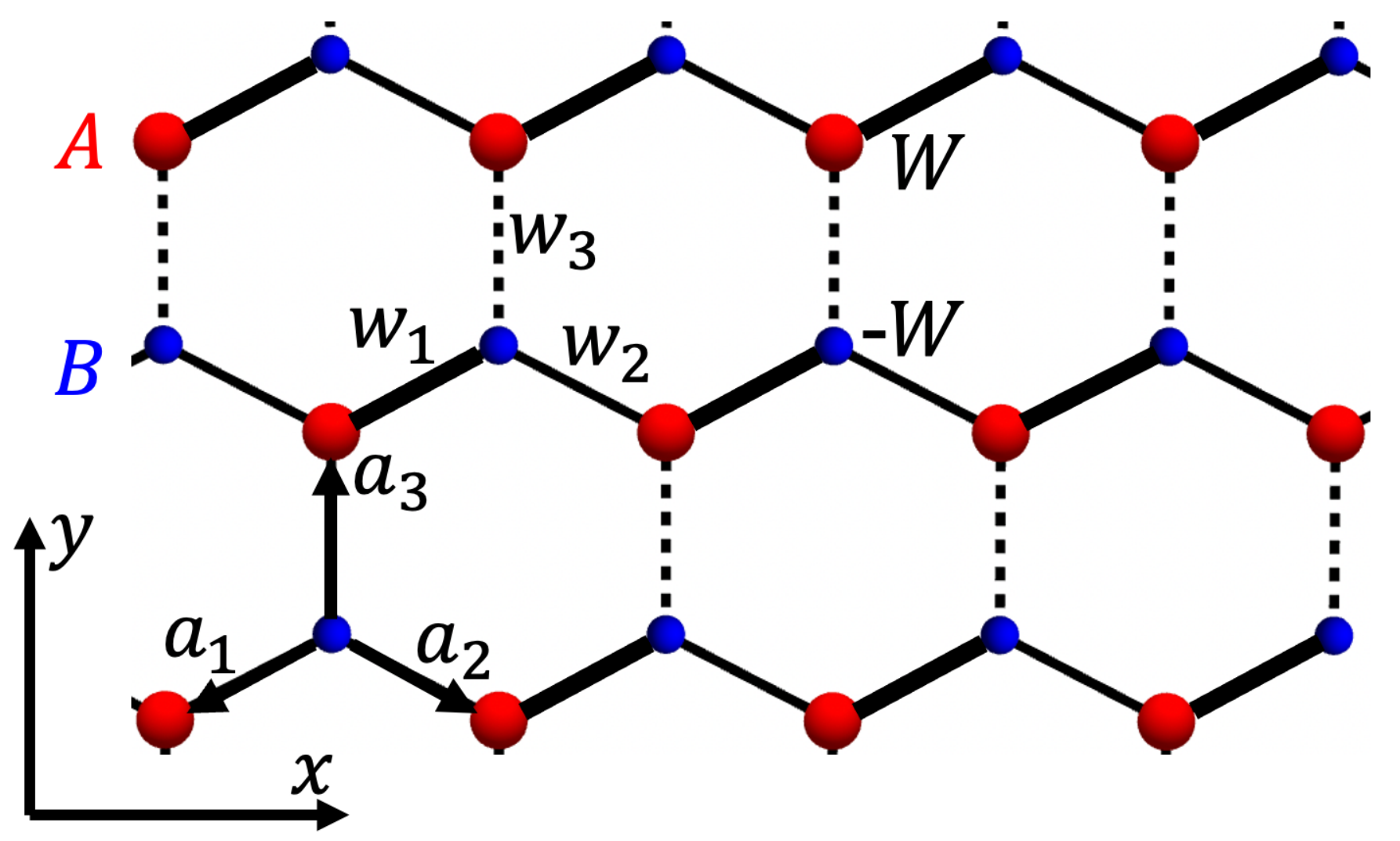}
\caption{Anisotropic honeycomb lattice with the hopping amplitudes $w_1$, $w_2$ and $w_3$ along the three nearest-neighbor bonds. The two sublattices $A$ and $B$ have the opposite on-site energies $W$ and $-W$. In the limit $w_3=0$, the system decouples into a set of zigzag chains along the $x$ axis.}
\label{fig:lattice}
\end{center}
\end{figure}

Motivated by the experimental realization of a shaking optical honeycomb lattice in Ref.~\cite{Jotzu2014}, we introduce an anisotropic 2D tight-binding lattice shown in Fig.~\ref{fig:lattice}. It has three tunneling amplitudes $w_1$, $w_2$, and $w_3$ along the nearest-neighbor vectors $\bm a_1$, $\bm a_2$, and $\bm a_3$ of the same length $a$. The two inequivalent sublattices A and B have different onsite energies $W$ and $-W$. This model breaks inversion symmetry but preserves time-reversal symmetry. 

In the 2D momentum representation where $\bm k = (k_x,k_y)$, the Hamiltonian is a $2\times2$ matrix expanded over the Pauli matrices $\bm \sigma = (\sigma_x,\sigma_y,\sigma_z)$
\begin{align}
& H(\bm k) = \bm \sigma \cdot \bm w(\bm k), \label{hamiltonian} \\
&  \bm w(\bm k)= \left[\sum_{j=1}^3 w_j\cos(\bm k\cdot\bm a_j), \sum_{j=1}^3 w_j\sin(\bm k\cdot\bm a_j),W\right].\nonumber
\end{align}
The Hamiltonian acts on the two-component wavefunction $u (\bm k) = [u_A(\bm k),u_B(\bm k)]^{\rm T}$ of the two sublattices.  The energy spectrum has the upper and lower bands with positive and negative eigenvalues 
\begin{align}
    \varepsilon_{m,n}(\bm k) = \pm |\bm w(\bm k)| 
    \label{general_spectrum}
\end{align}
labeled as $m$ and $n$, correspondingly.

In terms of the unit vector
\begin{equation}  \label{unit-w}
  \hat{\bm w}(\bm k) = \frac{\bm w(\bm k)}{|\bm w(\bm k)|},
\end{equation}
the Berry curvature of the lower band is
\begin{align}
    \Omega_{n,\gamma}(\bm k)   =\frac{\epsilon_{\alpha\beta\gamma}}{4} \,\hat{\bm w}(\bm k)\cdot\left[\frac{\partial\hat{\bm w}(\bm k)}{\partial k_\alpha}\times \frac{\partial\hat{\bm w}(\bm k)}{\partial k_\beta}\right]. \label{berry_curv}
\end{align}
Similarly, the symmetric product of the Berry connections appearing in Eq.~(\ref{two_band_sym}) has the form
\begin{align}
    &{\rm Re}\left[r^\alpha_{nm}(\bm k)r^\beta_{mn}(\bm k)\right] = \frac14\left[\frac{\partial\hat{\bm w}(\bm k)}{\partial k_\alpha}\cdot \frac{\partial\hat{\bm w}(\bm k)}{\partial k_\beta}\right], \label{connections_product}
\end{align}
which is shown in \ref{sec:berry_curv_appendix}.

We first study the model~(\ref{hamiltonian}) for $w_3 = 0$ in the limiting case of decoupled 1D zigzag chains in Sec.~\ref{sec:1d_limit}.  Then we proceed to the general 2D anisotropic honeycomb model in Sec.~\ref{sec:2D_limit}.

\section{One-dimensional limit: Decoupled chains}  \label{sec:1d_limit}
\begin{figure}
\begin{center}
\includegraphics[width=0.45\textwidth]{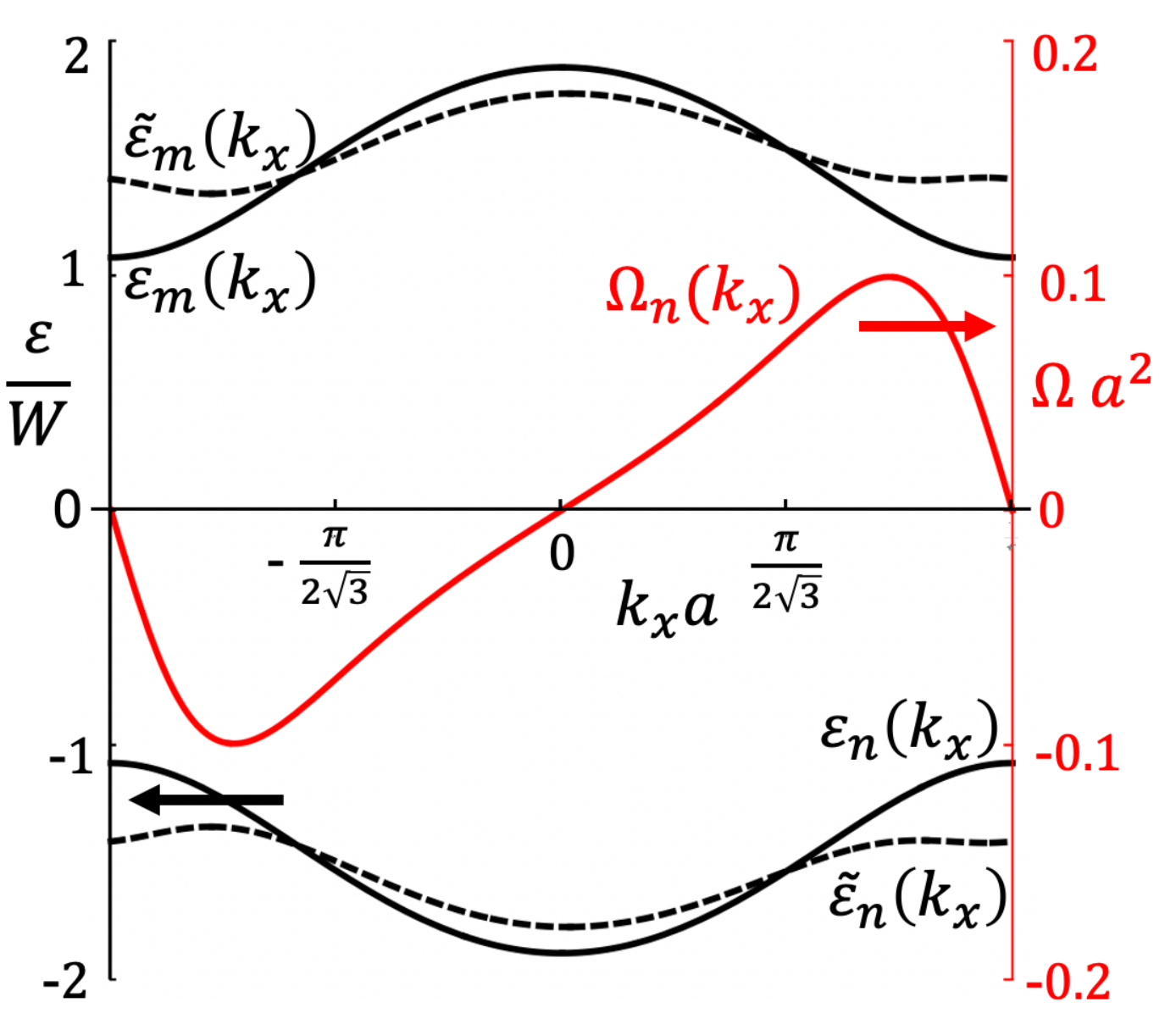}
\caption{The solid black and red curves show the bare energy spectrum (\ref{1d_spectrum}) and the Berry curvature (\ref{1d_berry}) for the decoupled 1D zigzag chains.  The dashed curves show the renormalized spectrum~(\ref{renormalized_spectrum}) modified by the terms (\ref{two_band_full}), (\ref{two_band_asym}), and (\ref{two_band_sym}) due to stirring of the lattice. The parameters are $W = w_1 = w_2/0.6 = \hbar\omega =  aF_+ /0.7$ and $w_3 = 0$.}
\label{fig:1Ddispersion}
\end{center}
\end{figure}

When one of the hopping amplitudes vanishes, e.g.,\ $w_3 = 0$, the 2D lattice shown in Fig.~\ref{fig:lattice} reduces to a set of decoupled 1D zigzag chains along the $x$ direction.  In this limit, it is known as the Rice-Mele model~\cite{Rice1982}. The energy spectrum~(\ref{general_spectrum}) becomes
\begin{align}
    &\varepsilon_{m,n}(\bm k) = \pm |\bm w(\bm k)| \label{1d_spectrum} \\
    & =\pm \sqrt{W^2+w_1^2+w_2^2+2\,w_1w_2\cos\bm k \cdot\left(\bm a_2- \bm a_1\right)}. \nonumber
\end{align}
It is shown by the solid black curve in Fig.~\ref{fig:1Ddispersion}.  Taking into account that the vector $(\bm a_2-\bm a_1)=\sqrt3 a \,\hat{\bm x}$ is along the chains in the $x$ direction, and so
\begin{align}
\bm k\cdot(\bm a_2-\bm a_1) = k_x\sqrt 3 a,
\end{align}
we note that the eigenenergies~(\ref{1d_spectrum}) depend only on the momentum  $k_x$ but not on $k_y$, because the chains are decoupled in the $y$ direction. Nevertheless, the Hamiltonian~(\ref{hamiltonian}) still depends on both components $k_x$ and $k_y$, so we obtain a nonzero Berry curvature for the lower band using Eq.~(\ref{berry_curv})
\begin{align}
  &\bm \Omega_n(\bm k) = \frac{W w_1w_2}{2\,|\bm w(\bm k)|^3}\,\, 
  [\bm a_1\times\bm a_2]\,\sin \bm k\cdot(\bm a_2-\bm a_1),  \label{1d_berry}
\end{align}
which depends only on the momentum $k_x$.  Substituting Eq.~(\ref{1d_berry}) into Eq.~(\ref{two_band_asym}) we find the antisymmetric energy shift for the 1D two-band model
\begin{align}  
  & \varepsilon_n^{(a)}(\bm k) = 
  -\frac{W w_1w_2}{2\hbar\omega |\bm w(\bm k)|}
  \frac{\bm h\cdot[\bm a_1\times\bm a_2]\, \sin\bm k\cdot(\bm a_2-\bm a_1)}
  {4|\bm w(\bm k)|^2-(\hbar\omega)^2} 
  \label{1D_asym}
\end{align}
The Berry curvature~(\ref{1d_berry}) is shown in Fig.~\ref{fig:1Ddispersion} by the solid red curve.  It is an odd function of $k_x$, because the Hamiltonian (\ref{hamiltonian}) has time-reversal symmetry, so it was called the {\it Berry curvature dipole} in Ref.~\cite{SodemannFu2015}.  The right-hand side of Eq.~(\ref{1d_berry}) is a product of the two momentum-dependent functions $1/|w(k_x)|^3$ and $\sin k_x \sqrt 3 a$. The former is maximal at the Brillouin zone edge $k_xa =  \pi/\sqrt 3$, whereas the latter at $k_xa = \pi/2\sqrt 3$. Thus the Berry curvature reaches maximum somewhere in the interval $\pi/2\sqrt 3 < k_xa <  \pi/\sqrt 3$. 

The dashed line in Fig.~\ref{fig:1Ddispersion} shows the energy dispersion~(\ref{renormalized_spectrum}) modified by the terms (\ref{two_band_full}), (\ref{two_band_sym}) and (\ref{1D_asym}) due to stirring of the lattice.  The dominant effect comes from the first term of the symmetric contribution~(\ref{two_band_sym}) proportional to the curvature $\partial^2\varepsilon/\partial k_\alpha \partial k_\beta$ and $1/\omega^2$.  As discussed in item (ii) below Eq.~(\ref{full_stark_shift}), this term leads to flattening of the band, which is visible in Fig.~\ref{fig:1Ddispersion}.

\begin{figure}
\begin{center}
\includegraphics[width=0.4\textwidth]{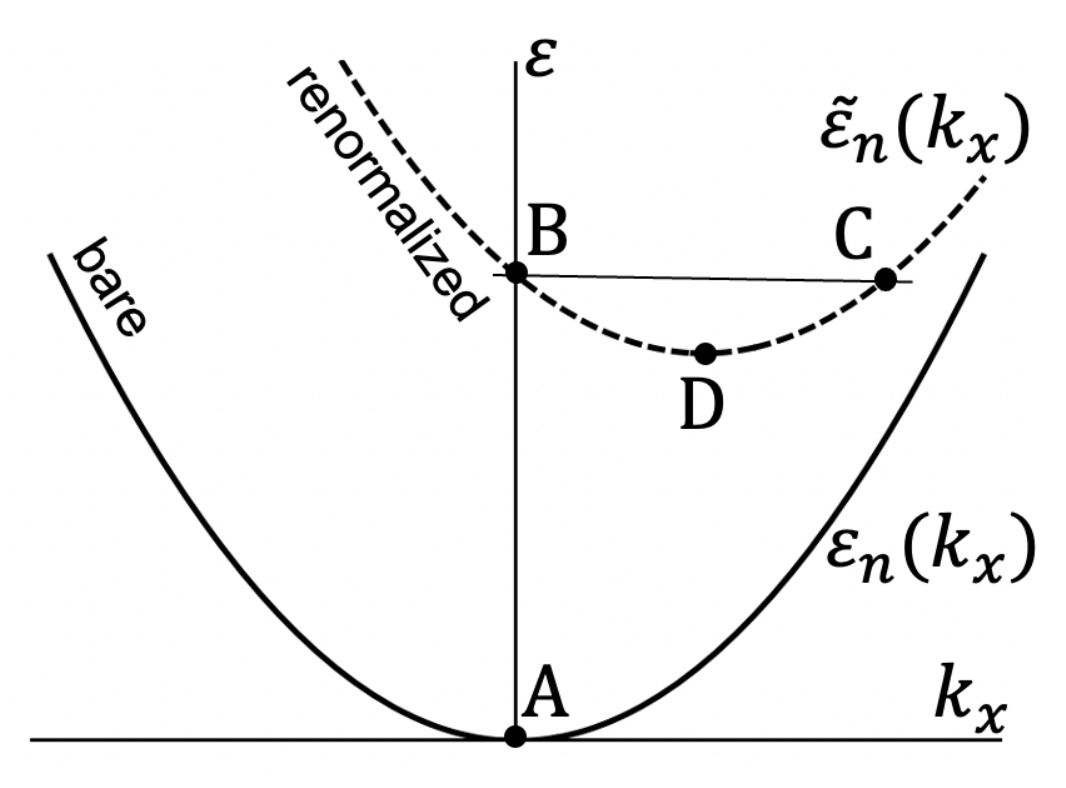}
\caption{The solid and dashed parabolas show the bare (\ref {bare_1d_dispersion}) and renormalized (\ref{renorm_1d_dispersion}) energy spectrum for the decoupled 1D zigzag chains.  The latter has a nonzero group velocity at $k_x = 0$ (point B).  In a spatially confined system, the atoms bounce between the states B and C with the opposite group velocities, but eventually relax to the stationary state D.}
\label{fig:renorm_dispersion}
\end{center}
\end{figure}

The antisymmetric term (\ref{1D_asym}) is odd in $\bm k$ and makes the energy dispersion asymmetric, as illustrated in Fig.~\ref{fig:renorm_dispersion} near $\bm k=0$.  Expansion of the bare spectrum~(\ref{1d_spectrum}) near its minimum yields
\begin{align}
    \varepsilon_n(\bm k) \approx \varepsilon_n(0) + \frac{\hbar^2\, k_x^2}{2\,M_\ast}, \label{bare_1d_dispersion}
\end{align}
where $\varepsilon_n(0) = -|\bm w(0)|=  -\sqrt{W^2+(w_1+w_2)^2}$ and $M_\ast = \hbar^2|\bm w(0)|/3w_1w_2\,a^2$.   In the presence of the antisymmetric term (\ref{1D_asym}), the modified energy dispersion (\ref{renormalized_spectrum}) acquires a linear-in-$\bm k$ term
\begin{align}
    \tilde\varepsilon_n(\bm k) \approx \tilde\varepsilon_n(0) + \frac{\hbar^2\, k_x^2}{2\,M_\ast}+\hbar\, \bm k \cdot\bm V, \label{renorm_1d_dispersion}
\end{align}
where the group velocity $\bm V$ at $\bm k=0$ is
\begin{align}
& \bm V = C_{12}\,  \, w_1w_2\,(\bm a_1-\bm a_2), \label{vel_x} \\
& C _{12}= \frac{W\,\bm h\cdot[\bm a_1\times\bm a_2]}
{2\,\hbar^2\omega|\bm w(0)|\left[4|\bm w(0)|^2-(\hbar\omega)^2\right]}. \nonumber
\end{align}
The bare dispersion~(\ref{bare_1d_dispersion}) is shown in Fig.~\ref{fig:renorm_dispersion} by the solid parabola centered at $k_x=0$, whereas the renormalized dispersion~(\ref{renorm_1d_dispersion}) by the dashed parabola,  shifted due to the linear-in-$\bm k$ term.

Here are some observations about the velocity~(\ref{vel_x}). 
(i) In order to move along the zigzag chains, the atoms have to traverse the nearest-neighbor bonds along the vectors $\bm a_1$ and $\bm a_2$ forming the chains.  Thus the velocity (\ref{vel_x}) is proportional to the product $w_1w_2$ of the corresponding hopping amplitudes. 
(ii) It is crucial that the vectors $\bm a_1$ and $\bm a_2$ are noncollinear to make a nonzero vector product in $C_{12}$. A straight 1D chain with $\bm a_1\|\bm a_2$ would not couple to the two components of the rotating force $\bm F$, and the velocity would vanish. 
(iii) The velocity~(\ref{vel_x}) changes sign when the direction of stirring is reversed as $\bm h \to -\bm h$.  
(iv) Following these considerations, the energy~(\ref{1D_asym}) and  velocity~(\ref{vel_x}) are derived in \ref{sec:perturbative} using a perturbation theory in coordinate space in the limit $w_1,w_2 \ll W$ without invoking the Berry curvature explicitly.

Let us discuss experimental manifestations of the renormalized dispersion~(\ref{renorm_1d_dispersion}) for cold bosonic atoms in the optical lattice.  At low temperature, the atoms occupy the lowest energy state of the bare dispersion at $k_x = 0$ marked as $A$ in Fig.~\ref{fig:renorm_dispersion}, where their group velocity is zero. Once the circular shaking is turned on, the energy dispersion (\ref{renorm_1d_dispersion}) becomes renormalized, as shown by the dashed parabola in Fig.~\ref{fig:renorm_dispersion}.  The atoms still have the same momentum $k_x=0$ (because the circular drive is spatially uniform), but their renormalized energy in now higher, marked as $B$ in Fig.~\ref{fig:renorm_dispersion}.  Importantly, the state $B$ now has a nonvanishing group velocity~(\ref{vel_x}), so the atoms start moving along the chains.  Since any optical lattice realized experimentally is spatially confined on some length scale, the atoms will eventually hit the edge of the system and reflect backward.  After reflection, they occupy the state $C$ in Fig.~\ref{fig:renorm_dispersion}, which is has the same energy as $B$, but the opposite velocity $-\bm V$.  The atoms will continue to oscillate back and forth along the chains by switching between the states $B$ and $C$, until they relax due to dissipation to the minimum D of the renormalized dispersion in Fig.~\ref{fig:1Ddispersion}, where the group velocity vanishes.  If the stirring is turned off after the atoms settle down at D, the energy dispersion will revert to the solid parabola, and the atoms will oscillate around the point A until they settle down at $k_x=0$.  Experimental measurements of the relaxation time can provide useful information about dissipation and thermalization in cold bosonic gases.  Also, by turning the circular stirring on and off for a limited time, the atoms can be transported over a fixed distance.  Moreover, in the 2D case discussed in the next Section, they can be transported in any direction by properly controlling the parameters of the optical lattice.

Technically, the dashed curve in Fig.~\ref{fig:renorm_dispersion} represents the Floquet quasi-energy, wheres the solid curve a conventional energy.  A question can be raised whether the system would relax to thermal equilibrium at $\tilde\varepsilon_n(\bm k)$ in the presence of steady drive.  It seems unlikely that the atoms would oscillate between B and C forever (as long as the drive is kept steady) and more likely that they would relax to the stationary point D.  The proposed experiment can give a definitive answer to this question and provide insight into thermalization of driven systems.  More discussion is given at the end of \ref{sec:kinetic}.

\section{Anisotropic 2D lattice} \label{sec:2D_limit}

\begin{figure}
\begin{center}
(a)\includegraphics[width=0.33\textwidth]{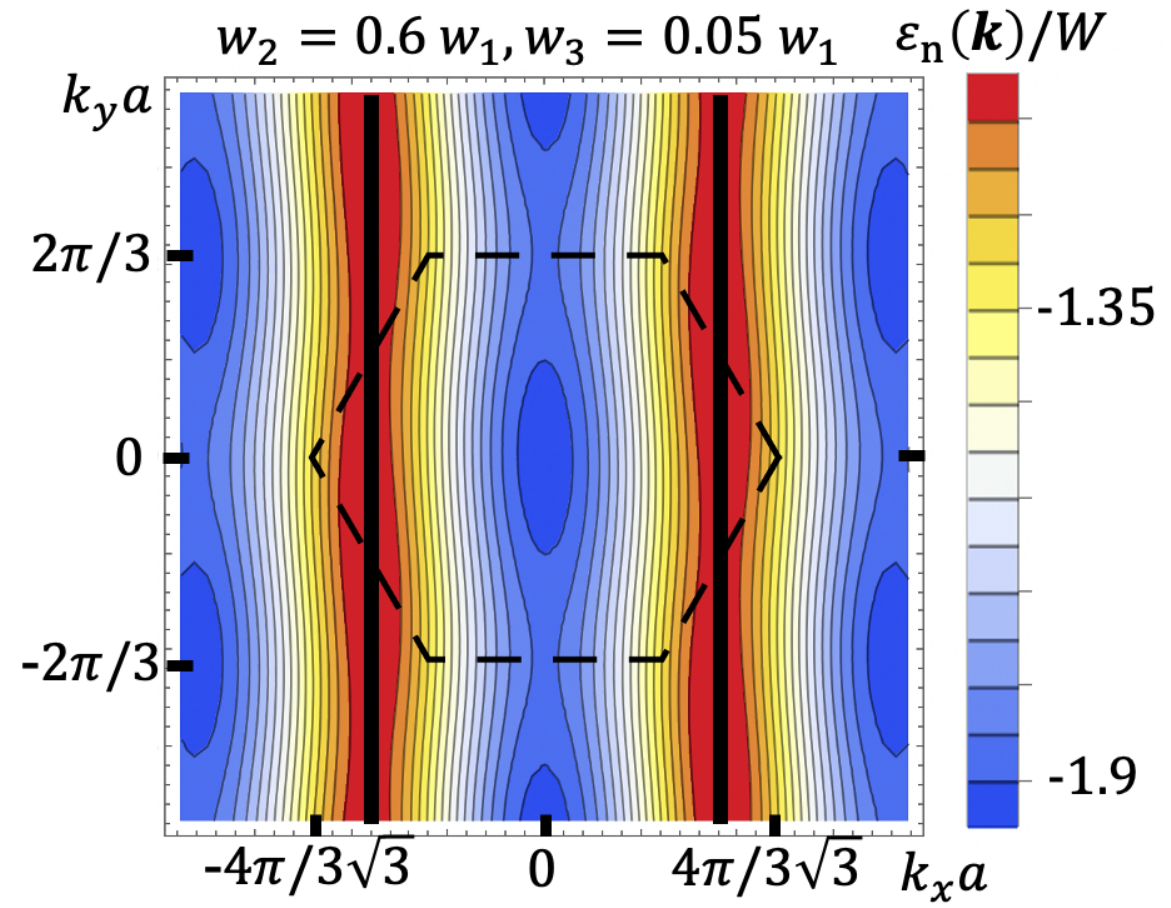} 
(b)\includegraphics[width=0.33\textwidth]{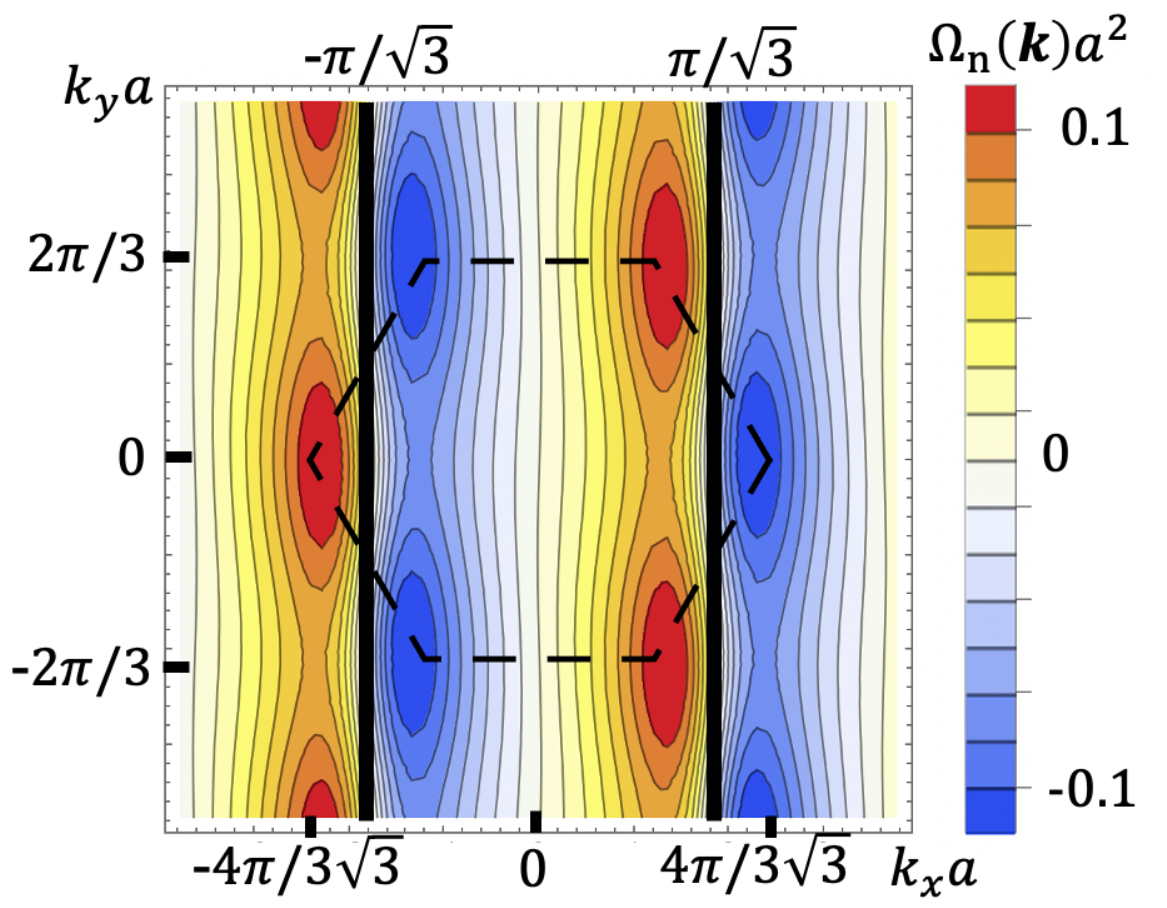} 
\caption{The bare energy dispersion (\ref{general_spectrum}) in panel (a) and the Berry curvature (\ref{berry_curv}) in panel (b) for a honeycomb lattice with highly anisotropic parameters $W = w_1 = w_2/0.6 = w_3/0.005$ corresponding to weakly coupled zigzag chains.  The dashed lines and the black vertical bars show the 2D and 1D Brillouin zone boundaries.}
\label{fig:dispersion1}
\end{center}
\end{figure}

\begin{figure}
\begin{center}
(a)\includegraphics[width=0.33\textwidth]{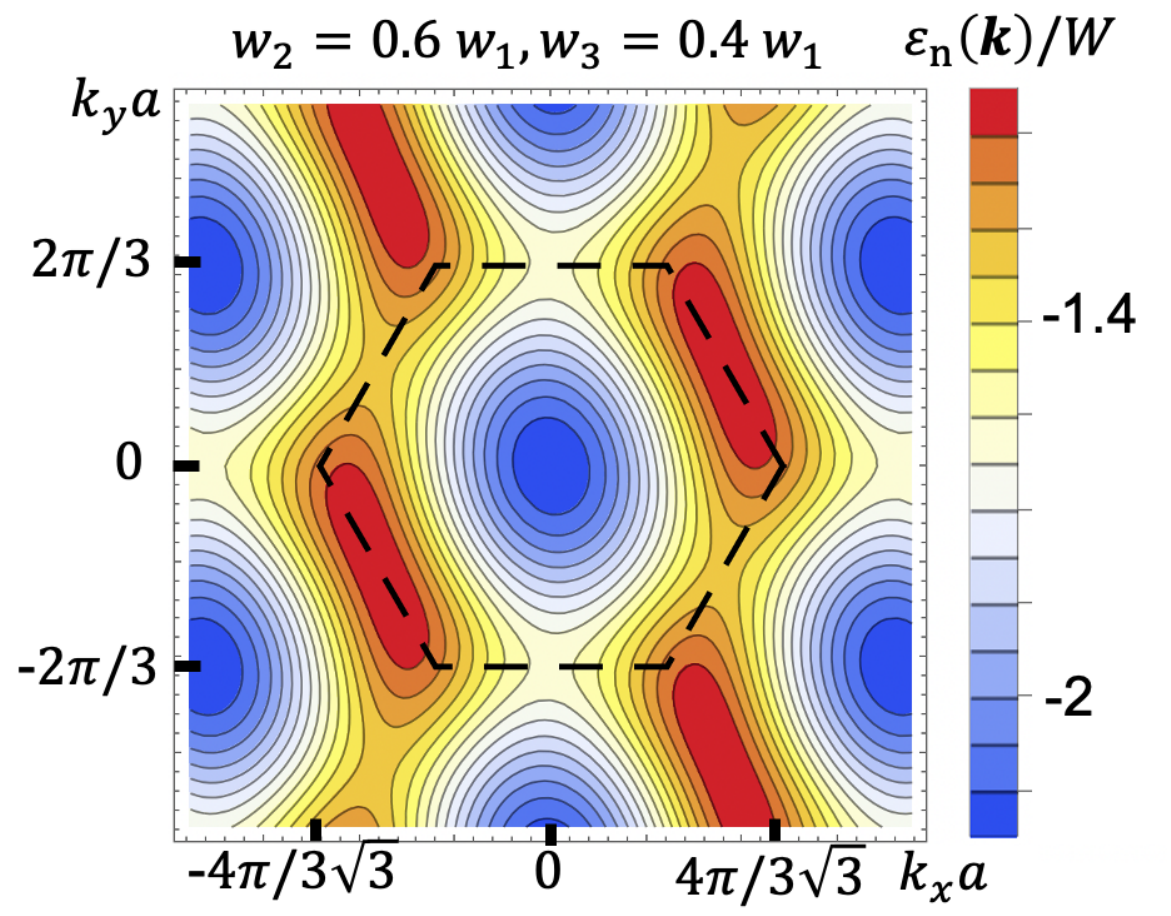} \\
(b)\includegraphics[width=0.33\textwidth]{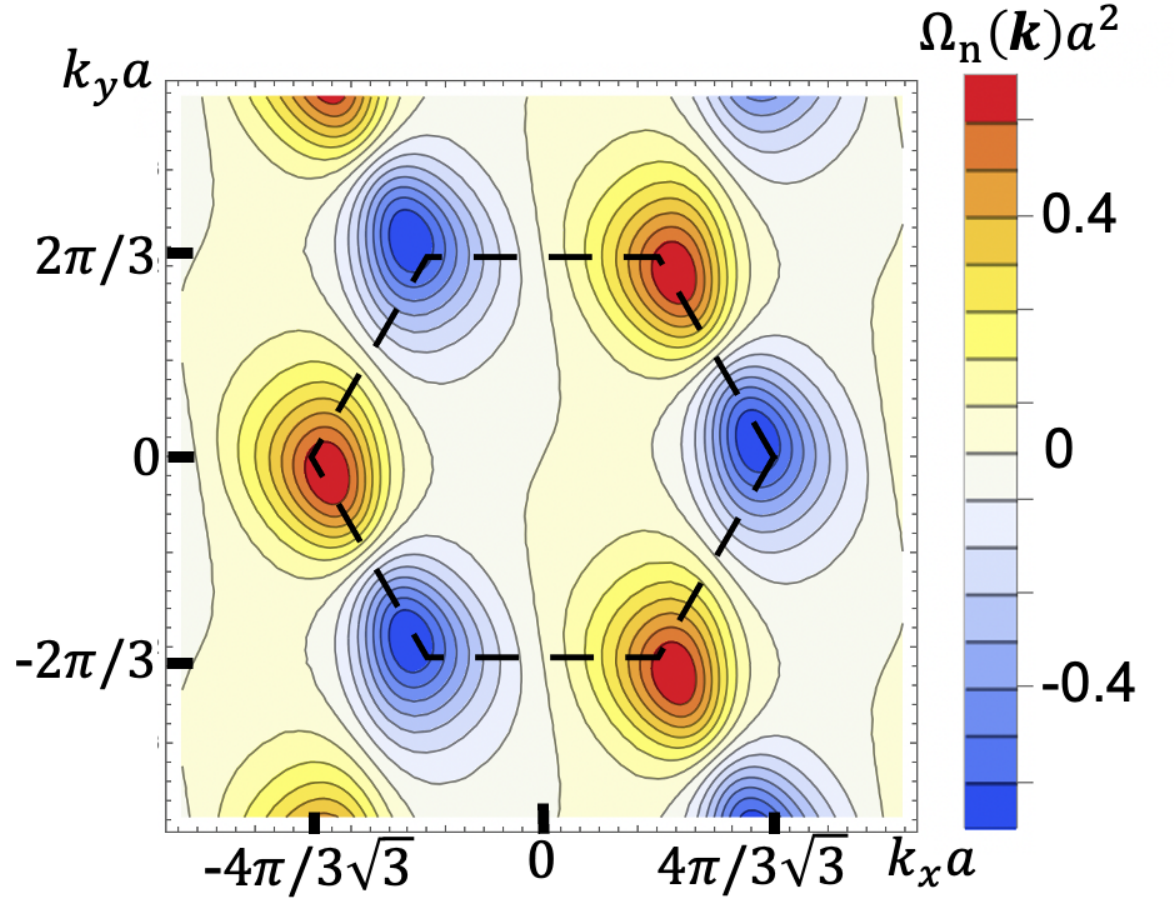}
\caption{Same as in Fig.~\ref{fig:dispersion1}, but for a moderately anisotropic lattice with the parameters $W = w_1 = w_2/0.6 = w_3/0.4$. }
\label{fig:dispersion2}
\end{center}
\end{figure}

Here we discuss a general anisotropic 2D honeycomb lattice, where all three hopping amplitudes $w_1$, $w_2$, and $w_3$ in Fig.~\ref{fig:lattice} and Eq.~(\ref{hamiltonian}) are nonzero.  To relate with the previous Section, we start with the highly anisotropic case $ w_1,\, w_2 \gg w_3$.  It corresponds to zigzag chains along $x$ weakly coupled by $w_3$ along $y$.  Figure~\ref{fig:dispersion1} shows that the corresponding energy spectrum and the Berry curvature are weakly modulated as functions of the momentum $k_y$.  The dashed lines and the black vertical bars show the 2D and 1D Brillouin zone boundaries, the latter corresponding to Fig.~\ref{fig:1Ddispersion}.  The case of a general anisotropic honeycomb lattice, where all hopping amplitudes of the same order, is shown in Fig.~\ref{fig:dispersion2}.  The Berry curvature has pronounced peaks near the corners of the Brillouin zone. 

Both in Figs.~\ref{fig:dispersion1}(b) and \ref{fig:dispersion2}(b), the Berry curvature has a nonzero gradient at $\bm k=0$.  Therefore, using Eq.~(\ref{vel}), we obtain the renormalized group velocity $\bm V$ at $\bm k = 0$, given by a generalization of Eq.~(\ref{vel_x})
\begin{align}
  \bm V & = C_{12}\, w_1w_2\,(\bm a_1-\bm a_2)  \nonumber \\ 
  & + C_{23}\, w_2w_3\,(\bm a_2-\bm a_3) \nonumber \\ 
  & + C_{31}\, w_3w_1\,(\bm a_3-\bm a_1) , \label{vel_general} \\
C _{ij} & = \frac{W\,\left[\bm h \cdot (\bm a_i \times \bm a_j)\right]}{2\,\hbar^2\omega\,|\bm w(0)|\left[4|\bm w(0)|^2-(\hbar\omega)^2\right]}. \nonumber
\end{align}
Equation~(\ref{vel_general}) has the following geometric interpretation. The honeycomb lattice in Fig.~\ref{fig:lattice} has three families of zigzag chains pointing along the vectors $\bm a_1-\bm a_2$, $\bm a_2-\bm a_3$, and $\bm a_3-\bm a_1$.  Equation~(\ref{vel_general}) implies that each chain family sustains a 1D current~(\ref{vel_x}) in the corresponding direction, and the total current is a vector sum of the three contributions. 

When all nearest-neighbor vectors have an equal magnitude $|\bm a_1|=|\bm a_2|= |\bm a_3| = a$, the prefactors $C_{ij}=C$ in Eq.~(\ref{vel_general}) are equal.  It is instructive to write the velocity~(\ref{vel_general}) in components parallel and perpendicular to the zigzag chains along $\bm a_1-\bm a_2$
\begin{align}
    \bm V &= \frac{C\,\left[2w_1w_2-w_3\left(w_1+w_2\right)\right]}{2} (\bm a_1-\bm a_2) \nonumber \\
    & + \frac{C\,3w_3\,(w_1-w_2)}2\,\bm a_3. \label{vel_general_b}
\end{align}
Some comments about Eq.~(\ref{vel_general_b}) are in order. 

(i) For the isotropic lattice, where all hopping amplitudes are equal $w_1=w_2=w_3$, the current vanishes because of rotational symmetry $C_3$. 

(ii)  If $w_1=w_2$, the velocity is purely along the zigzag direction $\bm V \parallel (\bm a_1-\bm a_2)\parallel \hat {\bm x}$. In this case, the lattice in Fig.~\ref{fig:lattice} is uniaxial and has a dipole $\bm P \parallel \hat{ \bm y}$ perpendicular to the zigzag chains, so the current direction is given by Eq.~(\ref{hXP})

(iii)\,If $w_3 = 2w_1w_2/(w_1+w_2)$, the current is along the armchair direction $\bm a_3$.  In general, the velocity $\bm V$ can be pointed in any direction by choosing the tunneling amplitudes appropriately.

\section{Conclusions} \label{sec:conclusions}

This paper proposes a generalization of the circular photogalvanic effect, much studied in solid-state physics, to neutral bosonic atoms in an optical lattice shifted periodically in time.  In the latter system, the force of inertia acting on bosons plays a similar role to the electric field acting on electrons.  The periodic drive produces a second-order correction to the energy dispersion \cite{PershogubaYakovenko2022}, whether for electrons in solids or bosons in a optical lattice, which has two important physical implications.

(i) The renormalized energy dispersion becomes flatter than the bare one, as shown in Fig.~\ref{fig:1Ddispersion} by the solid and dashed black curves.  It happens because the energy at a given momentum $\bm k$ averages out with the energies at the neighboring momenta due to the periodic drive.  This effect can be utilized to control the flatness of energy dispersion, which is a crucial parameter for emergence of strongly-correlated phases, e.g.,\ in the Moir\'e materials \cite{Andrei2021}. 

(ii) Circular stirring breaks the time-reversal symmetry and can produce an energy correction that is odd with respect to momentum, if the lattice breaks inversion symmetry.  
Consequently, the bosons, occupying the lowest energy state with $\bm k=0$, acquire a nonzero group velocity and produce a nonzero direct current.  This effect can be used to transport the neutral bosonic atoms in an optical lattice over desired distances and directions by turning the circular drive on and off at will.

These results are illustrated for an anisotropic 2D honeycomb optical lattice shown in Fig.~\ref{fig:lattice}, including the 1D limit of decoupled zigzag chains shown in Fig.~\ref{fig:geometry}, which is particularly illuminating pedagogically.  For a two-band model like this one, the odd energy shift discussed in item (ii) is expressed in terms of the Berry curvature of the lattice coupled to the helicity of the drive.

We focus on the case where the drive frequency is detuned so that there are no resonant interband transitions.  We find that the induced direct current in this case appears only due to modifications of the group velocity of the particles, as confirmed by the Hellmann-Feynman theorem in \ref{sec:Hellmann-Feynman} and the quantum kinetic equation in \ref{sec:kinetic}.  Thus, the induced current vanishes for an insulator with a completely filled band.  For a partially filled band, we conclude that the induced current is transient and vanishes after the particles relax to the thermal distribution $f[\tilde\varepsilon_n(\bm k)]$ corresponding to the modified energy dispersion $\tilde\varepsilon_n(\bm k)$, in which case the integral (\ref{current_group_velocity}) of the group velocity goes to zero.  Our conclusion is in agreement with Ref.~\cite{Belinicher1986} but in contrast with Refs.~\cite{SodemannFu2015,Nagaosa2022,Sodemann2022}.  While our mathematical equations essentially agree with those in Refs.~\cite{SodemannFu2015,Nagaosa2022}, the difference in physical conclusions is due to different assumptions toward what state the driven system relaxes in the presence of dissipation, as discussed at the end of \ref{sec:kinetic}.

Our arguments are illustrated in Fig.~\ref{fig:renorm_dispersion} for bosons in a 1D zigzag lattice, where the solid and dashed parabolas show the bare energy dispersion near $k_x$ and the renormalized one due to the drive.  The bosons are initially at point A and then get shifted to point B with a renormalized energy and a nonzero group velocity due to the drive.  For a spatially confined optical lattice, they will oscillate between points B and C upon reflections from the boundaries.  We argue that, due to dissipation, they will eventually relax to point D, where the group velocity is zero and the current vanishes.  An experimental verification of this prediction would help to understand equilibrization in driven systems.

\textit{Acknowledgments}. We thank J.~D.~Sau, D.~Pesin and L. Glazman for useful discussions at different stages of this work. SP was supported by the U.S.\ Department of Energy (DOE), Office of Science, Basic Energy Sciences (BES) under Award No. DE-SC0020221.

\bibliography{biblio}
\appendix
\section{Derivation of Eqs.~(\ref{berry_curv}) and (\ref{connections_product})} \label{sec:berry_curv_appendix}

For the $2\times 2$ Hamiltonian~(\ref{hamiltonian}), Eqs.~(\ref{berry_curv}) and (\ref{connections_product}) can be obtained in terms of the unit vector (\ref{unit-w}) without evaluating the eigenstates explicitly.  Let us denote the eigenstates as $|m\rangle$ and $|n\rangle$ for the positive and negative energies $\pm|\bm w(\bm k)|$ correspondingly.  We also introduce a concise notation 
\begin{equation}  \label{d_alpha}
    \partial^\alpha = \partial/\partial k_\alpha .
\end{equation}
Let us tackle the Berry curvature first.  We insert the identity operator $I=|n\rangle\langle n|+|m\rangle\langle m|$ into the definition (\ref{connections}) and (\ref{berry_curv_def}) of the Berry curvature
\begin{align*}
    \Omega_{n,\gamma} &= \epsilon_{\alpha\beta\gamma} \, i\,\langle \partial^\alpha n |  I|  \partial^\beta n \rangle  \\
    &= \epsilon_{\alpha\beta\gamma}  \,i\,\langle \partial^\alpha n | m\rangle \langle m|  \partial^\beta n \rangle .
\end{align*}
Here we used the antisymmetry with respect to interchange of the indices $\alpha \leftrightarrow \beta$ to drop the term with $| n\rangle \langle n |$ in the second line. The latter equation can be written using the trace
\begin{align}
    \Omega_{n,\gamma} &= \epsilon_{\alpha\beta\gamma}  \,i\,{\rm Tr} \left[\partial^\alpha (| n \rangle \langle n|) | m\rangle \langle m|  \partial^\beta (| n \rangle\langle n|) \right].
    \label{berry_interm}
\end{align}
The projection operators can be expressed using the Pauli matrices and the unit vector (\ref{unit-w})
\begin{equation}
  |m\rangle\langle m| = \frac{I+\bm\sigma\cdot\hat{\bm w}}{2}, \quad 
  |n\rangle\langle n| = \frac{I-\bm \sigma\cdot\hat{\bm w}}{2}.
  \label{projection}
\end{equation}

Substituting the projection operators~(\ref{projection}) into Eq.~(\ref{berry_interm}) and using the well-known identity for the Pauli matrices Tr$[\sigma_\alpha\sigma_\beta\sigma_\gamma] = 2\,i\,\epsilon_{\alpha\beta\gamma}$, we get
\begin{align*}
    \Omega_{n,\gamma}(\bm k)   =\frac{\epsilon_{\alpha\beta\gamma}}{4} \,\hat{\bm w}(\bm k)\cdot\left[\partial^\alpha\hat{\bm w}(\bm k)\times \partial^\beta\hat{\bm w}(\bm k)\right], 
\end{align*}
which is Eq.~(\ref{berry_curv}) of the main text.

Now, let us derive Eq.~(\ref{connections_product}) for the product of the Berry connections. We rewrite it using the trace
\begin{align}
     &{\rm Re}\,\left[r^\alpha_{nm}(\bm k)r^\beta_{mn}(\bm k)\right] =  \langle n| i \partial^\alpha m\rangle \langle m | i\partial^\beta n \rangle \nonumber \\
     & = -\frac12 {\rm Tr}\left[\partial^\alpha(| m \rangle\langle m|)\partial^\beta(| n \rangle\langle n|)\right]. \label{r_ar_b}
\end{align}
Substituting the projectors (\ref{projection}) and into Eq.~(\ref{r_ar_b}) using the identity Tr$[\sigma_\alpha\sigma_\beta] = 2\, \delta_{\alpha\beta}$, we obtain
\begin{align*}
     &{\rm Re}\,\left[r^\alpha_{nm}(\bm k)r^\beta_{mn}(\bm k)\right] =  \frac14\left[\partial^\alpha\hat{\bm w}(\bm k)\cdot \partial^\beta\hat{\bm w}(\bm k)\right],
\end{align*}
which is Eq.~(\ref{connections_product}) of the main text.

\section{Perturbative calculation in coordinate space} 
\label{sec:perturbative}
\begin{figure}
\begin{center}
\includegraphics[width=0.4\textwidth]{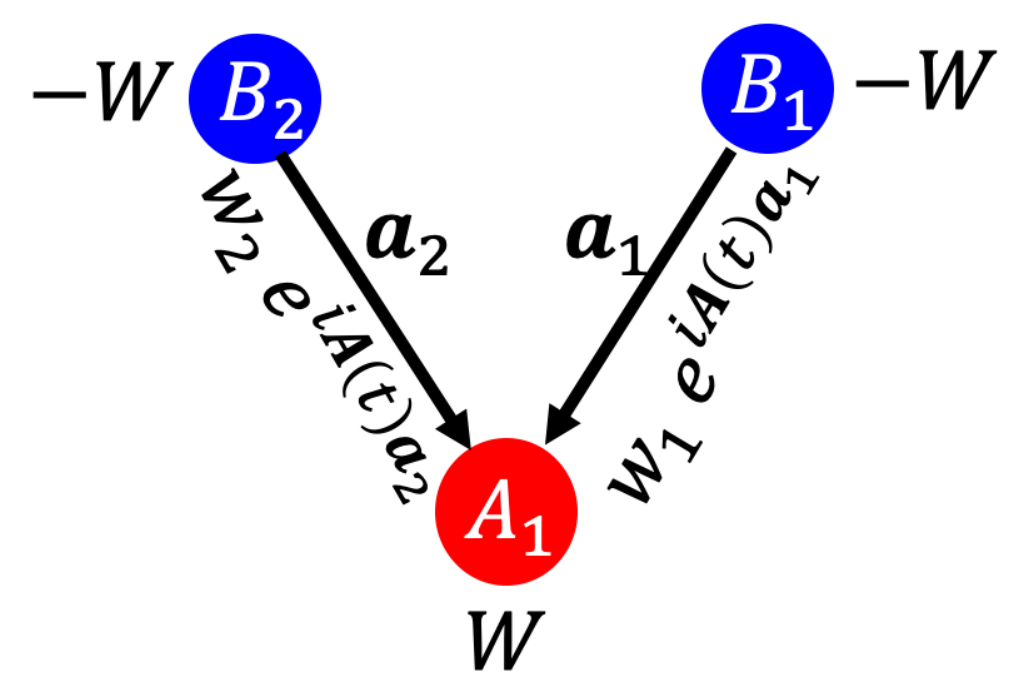}
\caption{Perturbation theory in the  nearest-neighbor hopping amplitudes $w_1$ and $w_2$ generates an effective coupling (\ref{w_eff}) between sites $B_1$ and $B_2$ with imaginary part.}
\label{fig:real_space}
\end{center}
\end{figure}

To clarify physical origin of the velocity~(\ref{vel_x}), here we derive it via a second-order perturbation theory in coordinate space without invoking the concept of the Berry curvature.  Let us consider three nearest sites from the whole lattice, labeled as $B_1$, $B_2$ and $A_1$ in Fig.~\ref{fig:real_space}.  Although there is no direct hopping amplitude between the next-nearest sites $B_1$ and $B_2$ in the Hamiltonian (\ref{hamiltonian}), an effective hopping amplitude $w_{\rm eff}$ connecting them appears in the second order of a perturbation theory in small $w_1$ and $w_2$.  We find that $w_{\rm eff}$ has an imaginary part due to circular stirring, which produces a nonzero velocity~(\ref{vel_x}). 

The unperturbed Hamiltonian for the three sites in the ``bra-ket'' notation is
\begin{align*}
    H_0 = W | A_1 \rangle \langle A_1 | - W  | B_1 \rangle \langle B_1 | - W  | B_2 \rangle \langle B_2 |.
\end{align*}
By analogy with the electric field, we introduce a vector potential $\bm {\mathcal A}(t)$ for the driving force via
\begin{equation}
  \bm F(t) = -\dot{\bm {\mathcal A}}(t)
\end{equation}
and use it for the harmonic force~(\ref{harmonic_expansion})
\begin{equation}  \label{A(t)}
\begin{aligned}
   & \bm {\mathcal A}(t) = \frac{\bm {\mathcal A}(\omega)e^{-i\omega t}+\bm {\mathcal A}(-\omega)e^{i\omega t}}2,\\
   & \bm {\mathcal A}(\omega) = \frac{\bm F(\omega)}{i\omega}.
\end{aligned}
\end{equation}
The vector potential produces nonzero phases for the hopping amplitudes in the tight-binding Hamiltonian connecting adjacent sites
\begin{align}
   H' & = w_1\, e^{i \bm {\mathcal A}(t)\cdot \bm a_1} | A_1 \rangle\langle B_1 |  \nonumber\\
   & + w_2\, e^{-i \bm {\mathcal A}(t)\cdot \bm a_2} | B_2 \rangle\langle A_1 | + {\rm H.c.},
   \label{H'}
\end{align}

Let us evaluate the evolution operator $U_I = \mathcal{T} \exp\left[\frac 1i \int^t dt'\, H'_{I}(t')\right]$ perturbatively in  $H'_{I}$ using the interaction ($I$) representation. We set $\hbar =1$ in the remaining Appendices.  We start with the first-order contribution to the evolution operator that moves a particle from site $B_1$ to $A_1$
\begin{align}
    &U_I^{(1)}(t) = \frac1i\int^t dt'\, H'_I(t') = \,| A_1 \rangle \langle B_1| \times\nonumber\\
    &\,\frac{w_1}{2i}\left[\frac{\mathcal A_\beta(\omega)\,e^{i(2W-\omega)t}}{2W-\omega}+ \frac{\mathcal A_\beta(-\omega)\,e^{i(2W+\omega)t}}{2W+\omega}\right]a_{1}^\beta \nonumber + \ldots,
\end{align}
where we expanded Eq.~(\ref{H'}) to the first order in $\bm {\mathcal A}$ and only retained the terms proportional to $\bm {\mathcal A}$. Next, we evaluate the second-order correction to the evolution operator and choose the process that moves the particle from site $A_1$ to $B_2$
\begin{align}
    &U_I^{(2)}(t) = \frac1i \int^t dt'\, H'_I(t')\, \frac1i \int^{t'} dt''\, H'_I(t'') \nonumber \\
    &=| B_2 \rangle \langle A_1| A_1 \rangle \langle B_1 | \frac 1i\int^t dt' \,\,w_{\rm eff} + \ldots \label{second_order}
\end{align}
This term generates the effective amplitude
\begin{align}
&\,w_{\rm eff}=-w_1w_2\frac{W\,{\rm Re}\left[\mathcal A_\alpha(-\omega)\mathcal A_\beta(\omega)\right]}{4W^2-\omega^2} a_1^\beta a_2^\alpha \nonumber\\
&\qquad-i\frac{w_1w_2}{2}\frac{\omega \,{\rm Im}\left[\mathcal A_\alpha(-\omega)\mathcal A_\beta(\omega)\right]}{4W^2-\omega^2} a_1^\beta a_2^\alpha \label{w_eff}
\end{align}
for transfer of the particle from site $B_1$ to site $B_2$ via the intermediate site $A_1$.  Notice that $w_{\rm eff}$ has an imaginary part that couples with ${\rm Im}\left[\mathcal A_\alpha(-\omega)\mathcal A_\beta(\omega)\right] = h^z \epsilon_{\alpha\beta z}/2\omega^2$, which is related to helicity $\bm h = h\, \hat{\bm z}$, so we get
\begin{align} \label{Im_w}
    {\rm Im}\left[w_{\rm eff}\right] 
    = \frac{w_1w_2\,\bm h\cdot[\bm a_1\times \bm a_2]}{4\,\omega(4W^2-\omega^2)}.
\end{align}

Now, let us consider the whole lattice with infinitely many sites. In momentum representation, the imaginary part of $w_{\rm eff}$ produces an odd-in-$\bm k$ contribution to energy dispersion
\begin{align} \label{sin_k}
    \varepsilon_{n}^{(a)}(\bm k) = 2\,{\rm Im}[w_{\rm eff}] \,\, \sin\bm k\cdot (\bm a_1 - \bm a_2).
\end{align}
Equations (\ref{Im_w}) and (\ref{sin_k}) agree with Eq.~(\ref{1D_asym}) in the limit $w_1,w_2\ll W$, where $|\bm w(\bm k)|\approx W$.

Correspondingly, the group velocity at $\bm k=0$
\begin{align}
    \bm V &= \left.\frac{\partial\varepsilon^{(a)}_{n}(\bm k)}{\partial \bm k}\right|_{\bm k=0} \\ 
    &= \frac{w_1w_2\,\bm h\cdot[\bm a_1 \times \bm a_2]}{2\,\omega(4W^2-\omega^2)} \, (\bm a_1 - \bm a_2). \nonumber
\end{align}
agrees with Eq.~(\ref{vel_x}) to the lowest order in $w_1w_2$.

\section{Hellmann-Feynman theorem for the Floquet states} 
\label{sec:Hellmann-Feynman}
There are two alternative methods for evaluation of the DC current induced by the driving force $\bm F(t)$. 
(i)  One method is to calculate the renormalized energy dispersion $\tilde\varepsilon_n(\bm k)$ and then express the current in terms of the group velocity $\partial\tilde\varepsilon_n(\bm k)/\partial \bm k$, as discussed in Sec.~\ref{sec:group-velocity}. 
(ii) Another approach is to calculate the expectation value of the current operator in the presence of a periodically oscillating force $\bm F(t)$.  In this Appendix, we prove a generalization of the Hellmann-Feynman theorem for the Floquet states and show that the two approaches (i) and (ii) give equivalent results.  

Let us consider the evolution operator for the period $T=2\pi/\omega$ of a time-periodic Hamiltonian $H$
\begin{align} \label{evolution_op}
    U(t+T,t) = \mathcal{T} \exp\left[\frac 1i \int_t^{t+T} dt' H(\bm k,t')\right].
\end{align} 
The Floquet states are the eigenstates of this evolution operator and satisfy the following equation
\begin{align} \label{floquet_def}
    e^{-i\,\tilde\varepsilon_n(\bm k) T} |n,\bm k,t\rangle = |n,\bm k,t+T\rangle  
    =  U(t+T,t) |n,\bm k,t\rangle, 
\end{align}
where $\tilde\varepsilon_n(\bm k)$ is the Floquet quasi-energy.  The wavefunctions in Eq.~(\ref{psi-k(t)}) are actually the Floquet states, and the renormalized energy $\tilde\varepsilon_n(\bm k)$ in Eq.~(\ref{renormalized_spectrum}) is the Floquet quasi-energy.

Let us differentiate the left- and right-hand sides of Eq.~(\ref{floquet_def}) over the momentum $\bm k$
\begin{align*}
   & - i\, T\,\frac{\partial\tilde\varepsilon_n(\bm k)}{\partial \bm k}\, e^{-i\,\tilde\varepsilon_n(\bm k) T} |n,\bm k,t\rangle +  e^{-i\,\tilde\varepsilon_n(\bm k) T} \frac{\partial}{\partial \bm k} |n,\bm k,t\rangle  \\
   & \quad = \frac{\partial U(t+T,t)}{\partial \bm k} |n,\bm k,t\rangle +  U(t+T,t) \frac{\partial}{\partial \bm k} |n,\bm k,t\rangle,
\end{align*}
and then multiply both sides by $\langle n,\bm k ,t+T|$
\begin{align}
   & - i\, T\,\frac{\partial\tilde\varepsilon_n(\bm k)}{\partial \bm k}\, +   \langle n,\bm k ,t|\frac{\partial}{\partial \bm k} |n,\bm k,t\rangle \nonumber  \\
   & \quad  = \langle n,\bm k ,t+T|\frac{\partial U(t+T,t)}{\partial \bm k} |n,\bm k,t\rangle \nonumber \\
   & \qquad +  \langle n,\bm k ,t| \frac{\partial}{\partial \bm k} |n,\bm k,t\rangle.
   \label{two-sides}
\end{align}
Upon cancellation of the identical terms in the left- and right-hand sides, Eq.~(\ref{two-sides}) reduces to
\begin{align} \label{interm_dUdt}
   & \frac{\partial\tilde\varepsilon_n(\bm k)}{\partial \bm k}  = \frac{i}{T}\langle n,\bm k ,t+T|\frac{\partial U(t+T,t)}{\partial \bm k} |n,\bm k,t\rangle. 
\end{align}
Differentiating the time-ordered exponential function in Eq.~(\ref{evolution_op}), we find
\begin{align} \label{identity_dUdt}
    &\frac{\partial U(t+T,t)}{\partial \bm k}  \\
    &=\frac{1}{i}\int_t^{t+T} dt'\, U(t+T,t')\, \frac{\partial H(\bm k,t')}{\partial \bm k} \, U(t',t). \nonumber
\end{align}
Using Eq.~(\ref{identity_dUdt}) in Eq.~(\ref{interm_dUdt}) and the definition of the evolution operator, we obtain the final result
\begin{align}
  \frac{\partial\tilde\varepsilon_n(\bm k)}{\partial \bm k} 
  =\int\limits_t^{t+T} \frac{dt'}T \langle n,\bm k ,t'|\frac{\partial H(\bm k,t')}{\partial \bm k}\,
  |n,\bm k,t'\rangle.  \label{Floquet_Feynman}
\end{align}
In the right-hand side of Eq.~(\ref{Floquet_Feynman}), the instantaneous quantum-mechanical expectation value is taken of the velocity operator ${\partial H(\bm k,t')}/{\partial \bm k}$ at time $t'$ with respect to the Floquet eigenstates $|n,\bm k,t'\rangle$ at the same $t'$.  This quantum expectation value generally depends on $t'$ and contains Fourier harmonics at zero frequency, $\omega$, $2\omega$, etc.  The subsequent integration over $t'$ eliminates higher harmonics and produces the DC value.  Thus Eq.~(\ref{Floquet_Feynman}) proves that the time-averaged expectation value of the velocity operator with respect to a Floquet state is identically equal to the group velocity of the corresponding Floquet quasi-energy.  Introducing the bar to represent time averaging and omitting $t'$ for shortness, Eq.~(\ref{Floquet_Feynman}) can be compactly written as 
\begin{align}
  \frac{\partial\tilde\varepsilon_n(\bm k)}{\partial \bm k}
  = \overline{\langle n,\bm k|\frac{\partial H(\bm k)}{\partial \bm k} \, |n,\bm k\rangle}.
  \label{Feynman_compact}
\end{align}
Equation (\ref{Feynman_compact}) is a generalization of the Hellmann-Feynman theorem for the Floquet states. 

In general, the Floquet quasi-energy $\tilde\varepsilon_n(\bm k)$ is ambiguously defined in Eq.~(\ref{floquet_def}) and can be shifted by a multiple of $\omega$, e.g., as $\tilde\varepsilon_n(\bm k)\to\tilde\varepsilon_n(\bm k)+\omega$.  But this ambiguity does not affect Eq.~(\ref{Feynman_compact}), because the derivative of $\omega$ vanishes in the left-hand side.

We should clarify that Eq.~(\ref{Feynman_compact}) is not suitable for obtaining a linear response at the frequency $\omega$, which is filtered out by time averaging.  Instead, a linear current response can be obtained by appropriate differentiation of Eq.~(\ref{full_stark_shift}) with respect to $\bm{\mathcal A}(\omega)$ defined in Eq.~(\ref{A(t)}), as discussed in Ref.~\cite{PershogubaYakovenko2022}.  This procedure gives the Hall current at the frequency $\omega$ and reproduces the quantum Hall effect for a Chern insulator in the limit $\omega\to0$ due to the second term in Eq.~(\ref{full_stark_shift}).

Equation (\ref{Feynman_compact}) is exact and valid to any perturbative order in the driving force $\bm F$.  To be sure, we explicitly evaluated the Floquet eigenstates to the second order in $\bm F$ using the formalism of Ref.~\cite{Bauer2020} for a two-band model and confirmed Eq.~(\ref{Floquet_Feynman}) to that order. This long calculation is not presented here because of the general proof given above.  Instead, in the following Appendix, we present an alternative derivation of the DC current using the quantum kinetic equation~\cite{Sipe2000}, which is often used in the literature to evaluate the non-linear response. 

\section{Calculation of current using the quantum kinetic equation} \label{sec:kinetic}

In this Appendix, we derive the DC current $\bm j$ from the quantum kinetic equation, which is a popular approach in the literature \cite{Sipe2000}.  In the end, we show that the result is equivalent to Eq.~(\ref{current_group_velocity}) expressed in terms of a modified group velocity.

For a general multiband model, where the indices $n$ and $m$ label the bands, the quantum kinetic equation describes the time evolution of the density matrix $f_{nm}$ under the influence of the driving force $F_\alpha(t)$
\begin{align}
  \dot f_{nm} + i \,\varepsilon_{nm}\, f_{nm} = F_\alpha(t) \left(\partial^\alpha f_{nm} - i[r^\alpha,f]_{nm}\right). \label{kinetic_general}
\end{align}
Here we set $\hbar=1$ and use the concise notation (\ref{d_alpha}).  In the last term, the square brackets represent a commutator, and $r^\alpha$ denotes the matrix of the Berry connections (\ref{connections}).  The Berry connections $r^\alpha(\bm k)$ and the band energies differences $\varepsilon_{nm}(\bm k)=\varepsilon_n(\bm k)-\varepsilon_m(\bm k)$ depend on the momentum $\bm k$, whereas the density matrix $f(t,\bm k)$ also depends on  time $t$.  We omit these arguments to shorten notation.  Equation (\ref{kinetic_general}) is written in the dissipationless limit and does not include relaxation, so it is ultimately equivalent to the time-dependent Schr\"odinger's equation.  Dissipation and relaxation are briefly discussed at the end of the Appendix.  A derivation of Eq.~(\ref{kinetic_general}) is given in Ref.~\cite{Sipe2000}.

Next we focus on a two-band model and reserve the labels $n$ and $m$ specifically for the lower and the upper bands, respectively. Then, the density matrix and the matrix of Berry connections are $2\times2$
\begin{align}
    f = \left(\begin{array}{cc}
     f_{mm} & f_{mn} \\ f_{nm} & f_{nn}
    \end{array}\right), \quad 
    r^\alpha = \left(\begin{array}{cc}
     r^\alpha_{mm} & r^\alpha_{mn} \\ r^\alpha_{nm} & r^\alpha_{nn}
    \end{array}\right).
\end{align}
The density matrix has two real diagonal matrix elements, which we replace by the traces 
\begin{equation}
\begin{aligned}
  & \bar f= {\rm Tr}\, f = f_{mm}+f_{nn}, \\
  & f_z = {\rm Tr}\, \sigma_zf = f_{mm}-f_{nn},
\end{aligned} \label{fzi}
\end{equation}
as well as the complex off-diagonal matrix elements $f_{mn} = f^\ast_{nm}$ related by complex conjugation.  Then Eq.~(\ref{kinetic_general}) reduces to three differential equations
\begin{align}
    & \dot{\bar f}_i = F_\alpha(t) \,\partial^\alpha \bar f, \label{fi} \\
    & \dot f_z = F_\alpha(t) \left[\partial^\alpha f_z + 4 \,{\rm Im}\left(r^\alpha_{mn}f_{nm}\right)\right], \label{fz} \\
    & \dot f_{nm} + i\,\varepsilon_{nm}\,f_{nm} = F_{\alpha}(t) \left[D^\alpha_{nm}f_{nm}-i\,r^\alpha_{nm}f_z\right], \label{fbt}
\end{align}
where 
\begin{align}
    D^\alpha_{nm} = \partial^\alpha-i(r^\alpha_{nn}-r^\alpha_{mm})
\end{align}
is a gauge-invariant long derivative. Summation is implied over the spatial index $\alpha$ in Eqs.~(\ref{fi})--(\ref{fbt}), but not over the band indices $m$ and $n$.

Equation (\ref{fi}) for $\bar f$ decouples from Eqs.~(\ref{fz}) and (\ref{fbt}).  Its solution has the form $\bar f(t,\bm k) = \bar f^{(0)}[\bm k+\delta {\bm k}(t)]$, where $\tilde {\bm k}(t)  = \bm k+\delta {\bm k}(t)$ is the shifted momentum due to the driving force in Eq.~(\ref{Newton}), and $\bar f^{(0)}(\bm k)$ is the initial distribution. It reflects the fact that the total probability of finding the particles in the two bands is conserved over time. Thus, $\bar f$ is irrelevant for the effect under consideration. 

The remaining two equations~(\ref{fz}) and (\ref{fbt}) can be solved iteratively as
\begin{equation}  \label{iterations}
  f = f^{(0)} + f^{(1)} + f^{(2)} + \ldots
\end{equation}
using the driving force as a perturbation
\begin{align}
 F_\alpha(t) = \frac{F_\alpha \,e^{-i\omega t+\delta t} + F^\ast_\alpha\, e^{i\omega t+\delta t}}{2}.
\end{align}
The small parameter $\delta>0$ simulates the adiabatic turn on the driving force from $t = -\infty$ to some finite $t$. We use the following initial conditions 
\begin{align}
    & f_z(t,\bm k)|_{t=-\infty} = f_z^{(0)}(\bm k), \label{init_cond_z} \\
    & f_{nm}(t,\bm k)|_{t=-\infty} = 0, \label{init_cond_nm}
\end{align}
where the superscript $(0)$ indicates the initial distribution function in momentum space.

{\it First-order correction}. Using the initial conditions (\ref{init_cond_z}) and (\ref{init_cond_nm}) in Eqs.~(\ref{fz}) and (\ref{fbt}), we obtain equations for the first-order corrections
\begin{equation*}
    \begin{aligned}
    \dot f^{(1)}_z &= F_\alpha(t) \,\partial^\alpha f^{(0)}_z,  \\
    \dot f^{(1)}_{nm} &+ i\,\varepsilon_{nm}\,f^{(1)}_{nm} = -i F_{\alpha}(t)\, r^\alpha_{nm}\,f^{(0)}_z. 
\end{aligned}
\end{equation*}
Integrating these equation over time, we find the first-order correction to the density matrix
\begin{align}
    f_z^{(1)} &= \frac{i}{2} \left[\frac{F_\alpha\,e^{-i\omega t+t\delta}}{\omega+i\delta}+\frac{F^\ast_\alpha\,e^{i\omega t+t\delta}}{-\omega+i\delta}\right] \partial^\alpha f_z^{(0)}, \label{first_order_z} \\
    f_{nm}^{(1)} &= \frac 12 \left[\frac{F_\alpha\, e^{-i\omega t+t\delta}}{\varepsilon_{nm}+\omega+i\delta}+\frac{F^\ast_\alpha\, e^{i\omega t+t\delta}}{\varepsilon_{nm}-\omega+i\delta}\right]r^\alpha_{nm}\, f_z^{(0)}.
    \label{first_order_nm}
\end{align} 
The effect of the driving force is two-fold.  In Eq.~(\ref{first_order_z}), it causes a shift of the diagonal distribution function $f_z^{(1)}$ in momentum space consistent with Newton's law (\ref{Newton}).  In Eq.~(\ref{first_order_nm}), it generates the off-diagonal component $f_{nm}^{(1)}$
representing a coherent superposition of the lower and upper states.

{\it Second-order correction}. Now we substitute the first-order corrections~(\ref{first_order_z}) and (\ref{first_order_nm}) into the right-hand sides of Eqs.~(\ref{fz}) and (\ref{fbt}) and obtain  equations on the second-order corrections:
\begin{equation*}
    \begin{aligned}
        \dot f^{(2)}_z &= F_\beta(t) \left[ \partial^\beta f^{(1)}_z + 4\,{\rm Im} \left(r^\beta_{mn}f^{(1)}_{nm}\right)\right], \\
        \dot f^{(2)}_{nm} &+ i\,\varepsilon_{nm} f^{(2)}_{nm} = F_\beta(t) \left[D^\beta_{nm}\, f^{(1)}_{nm} - i\, r^\beta_{nm}\,f_z^{(1)}\right].
    \end{aligned}
\end{equation*}
Since we are only interested in the DC current, we drop the terms oscillating at the double frequency $2\omega$ in the right-hand side and only keep the antisymmetric terms related to the helicity (\ref{antisymmetric})
    \begin{align}
        &\dot f^{(2)}_z = - \frac{2\,(\bm h\cdot\bm \Omega_n)\,\varepsilon_{mn}\, \omega\, \delta\, e^{2\delta t}}{\left[(\varepsilon_{mn}+\omega)^2+\delta^2\right]\left[(\varepsilon_{mn}-\omega)^2+\delta^2\right]}\,f_z^{(0)}, \label{cor2_z} \\
        &\dot f^{(2)}_{nm} + i\,\varepsilon_{nm} f^{(2)}_{nm}= \frac{ih^\gamma\epsilon_{\alpha\beta\gamma}}{4} \nonumber \\ 
        & \times \left[D^\beta_{nm} \left(\frac{\omega\, r^\alpha_{nm}\, f^{(0)}_z}{(\varepsilon_{mn}+i\delta)^2-\omega^2}\right)-\frac{\omega\, r^\beta_{nm}\,\partial^\alpha f_z^{(0)}}{\omega^2+\delta^2}\right] e^{2\delta t}.
        \label{cor2_nm}
    \end{align} 
Here we used Eq.~(\ref{Berry_relation}) for the Berry curvature $\bm\Omega_n$. 

For subgap frequencies $\omega<|\varepsilon_{mn}|$, the small parameter $\delta$ can be dropped in the denominators of Eqs.~(\ref{cor2_z}) and (\ref{cor2_nm}).  However, it is important to keep  $\delta\,e^{2\delta t}$ in the numerator of Eq.~(\ref{cor2_z}).  Only after integration over time, the limit $\delta\to0$ can be taken, which gives a nonzero result
\begin{equation}
  \int_{-\infty}^t \delta\, e^{2\delta t'} \, dt' = \frac12\, e^{2\delta t} \to \frac12
  \quad{\rm at}\quad\delta\to0.
\end{equation}
In contrast, $\delta$ can be omitted right away for integration of Eq.~(\ref{cor2_nm}) over time.  Thus we obtain the second-order corrections to the density matrix
    \begin{align}
        &f^{(2)}_z = \frac{\bm h\cdot \bm \Omega_n}{2}\,\frac{\partial}{\partial \varepsilon_{mn}}\left[\frac{\omega}{\varepsilon_{mn}^2-\omega^2}\right] f_z^{(0)},
        \label{second_order_f_z} \\
        &f^{(2)}_{nm} = \frac{h^\gamma\epsilon_{\alpha\beta\gamma}}{4\,\varepsilon_{nm}} \left[D^\beta_{nm} \left(\frac{\omega\, r^\alpha_{nm}\, f^{(0)}_z}{\varepsilon_{mn}^2-\omega^2}\right)-\frac{r^\beta_{nm}\,\partial^\alpha f_z^{(0)}}{\omega}\right] . \label{second_order_f_nm}
    \end{align} 

{\it Current density}. Now we evaluate the current density produced by the driving force
\begin{align}
    &j^\eta = \int_{\bm k}\,{\rm Tr}\left[f\,v^\eta \right] = \label{trace}\\ 
    & \int_{\bm k}\,\left[\frac 12 \bar f\, \partial^\eta \bar\varepsilon_{nm} + \frac 12 f_z\, \partial^\eta\varepsilon_{mn} + 2\,\varepsilon_{nm}\,{\rm Im}\left(r^\eta_{mn} f_{nm}\right)\right]. \nonumber
\end{align}
Here we denote $\bar\varepsilon_{nm}=\varepsilon_n+\varepsilon_m$ and the momentum-space integral
as $\int_{\bm k} \equiv \int d^d\bm k/(2\pi)^d$.  The first term with $\bar f$ in Eq.~(\ref{trace}) does not contribute to the helicity-induced current, so we drop it.  The second term gives a contribution to the current due to differences in the occupations and group velocities of the upper and lower bands.  The last contribution is from the interband terms in the density matrix. 

Substituting the second-order corrections~(\ref{second_order_f_z}) and (\ref{second_order_f_nm}) into Eq.~(\ref{trace}), we obtain
\begin{align*}
    j^\eta &= \int_{\bm k} \frac{\bm h\cdot \bm \Omega_n}{4}\,\partial^\eta\left[\frac{\omega}{\varepsilon_{mn}^2-\omega^2}\right] f_z^{(0)} \\
    &\quad+ {\rm Im} \int_{\bm k}\,\frac{h^\gamma \,\epsilon_{\alpha\beta\gamma}\, r^\eta_{mn}}{2}\, D^\beta_{nm} \left(\frac{\omega\,r^\alpha_{nm}f_z^{(0)}}{\varepsilon_{mn}^2-\omega^2}\right) \\
    &\quad - {\rm Im} \int_{\bm k}\,\frac{h^\gamma \,\epsilon_{\alpha\beta\gamma}\, r^\eta_{mn}r^\beta_{nm}\,\partial^\alpha f_z^{(0)}}{2\,\omega}.
\end{align*}
We integrate the second and third terms by parts
\begin{align}
    j^\eta &= \int_{\bm k} \frac{\bm h\cdot \bm \Omega_n}{4}\,\partial^\eta\left[\frac{\omega}{\varepsilon_{mn}^2-\omega^2}\right] f_z^{(0)} 
    \nonumber \\
    &\quad- {\rm Im} \int_{\bm k}\,\frac{h^\gamma \,\epsilon_{\alpha\beta\gamma}\,\left(r^\alpha_{nm} D^\beta_{mn}r^\eta_{mn} \right)}{2}\,  \frac{\omega}{\varepsilon_{mn}^2-\omega^2}f_z^{(0)}
    \nonumber \\
    &\quad+ {\rm Im} \int_{\bm k}\,\frac{h^\gamma \,\epsilon_{\alpha\beta\gamma}\, \,\partial^\alpha\left(r^\eta_{mn}r^\beta_{nm}\right)}{2\,\omega}f_z^{(0)}.
    \label{integral_interm}
\end{align}
The coefficients in the last two terms can be expressed in terms of the Berry curvature of the lower band using the identity for a two-band model 
\begin{align*}
     &\epsilon_{\alpha\beta\gamma}{\rm Im} \left(r^\alpha_{nm} D^\beta_{mn}r^\eta_{mn}\right) = \epsilon_{\alpha\beta\gamma} {\rm Im} \left(r^\alpha_{nm} D^\eta_{mn}r^\beta_{mn}\right) \\
     &= \frac {\epsilon_{\alpha\beta\gamma}}2\partial^\eta{\rm Im} \left(r^\alpha_{nm} r^\beta_{mn}\right) 
     = -\frac 12 \,\partial^\eta\,\Omega_{n,\gamma}.
\end{align*}
Then Eq.~(\ref{integral_interm}) becomes
\begin{align}
    j^\eta &= \int_{\bm k} \frac{\bm h\cdot \bm \Omega_n}{4}\,\partial^\eta\left[\frac{\omega}{\varepsilon_{mn}^2-\omega^2}\right] \, f_z^{(0)} 
    \nonumber \\
    &\qquad+  \int_{\bm k}\,\partial^\eta\left[\frac{\bm h\cdot \bm \Omega_n}{4}\right]\,  \frac{\omega}{\varepsilon_{mn}^2-\omega^2}\,f_z^{(0)}
    \nonumber \\
    &\qquad+  \int_{\bm k}\,\partial^\eta\left[\frac{\bm h\cdot \bm\Omega_n}{4\,\omega}\right]\,f_z^{(0)} 
    \nonumber \\   
    &= \int_{\bm k} \partial^\eta\left[\frac{\bm h\cdot \bm \Omega_n}{4}\,\frac{\omega}{\varepsilon_{mn}^2-\omega^2}+\frac{\bm h\cdot \bm\Omega_n}{4\,\omega}\right] \, f_z^{(0)} 
    \nonumber \\    
    &= \int_{\bm k} \partial^\eta\left[\frac{\bm h\cdot \bm \Omega_n}{4\,\omega}\,\frac{\varepsilon_{mn}^2}{\varepsilon_{mn}^2-\omega^2}\right] \, f_z^{(0)}.
    \label{j_current}
    \end{align} 
The structure of Eq.~(\ref{j_current}) for the current $\bm j$ is the same as in Eq.~(\ref{current_group_velocity}) and agrees with Ref.~\cite{SodemannFu2015}.  It involves a product of the group velocity due to the modified energy dispersion (\ref{two_band_asym}) and the occupation function $f_z$.  The latter represents the  difference (\ref{fzi}) of the upper and lower occupation numbers and appears in Eq.~(\ref{j_current}) because the Berry curvature has opposite signs in the two bands.

Equation (\ref{j_current}) supports the claim made in Eq.~(\ref{current_density}) that the DC current $\bm j$ induced to the second order in $\bm F(\omega)$ is determined solely by the modified group velocity, and there is no ``anomalous'' velocity. (This is in contrast to the linear response at frequency $\omega$, where is anomalous velocity in addition to the group velocity.)  As a consequence, the induced current $\bm j$ vanishes for an insulator, where the lower band is fully occupied and the upper band is empty, so that $f_z^{(0)}=-1$, and Eq.~(\ref{j_current}) becomes an integral of a full derivative.

Equation (\ref{j_current}) contains the initial occupation function $f_z^{(0)}(\bm k)$, which can be written as $f_z^{(0)}[\varepsilon(\bm k)]$, when the initial distribution is thermal.  For bosons at low temperature, it is approximately the delta-function $f^{(0)}(\bm k)\propto\delta^{(d)}(\bm k)$.  Substituting it into Eq.~(\ref{j_current}), we obtain a nonzero induced current proportional to the velocity $\bm V$ given by Eq.~(\ref{vel}).  However, there is a physical question whether this induced current would persist forever (as long as the drive is present), or is transient and will eventually relax to zero.  We argue for the latter conclusion.

For the setup shown in Fig.~\ref{fig:renorm_dispersion}, Eq.~(\ref{j_current}) can be interpreted in the following way.  Intially, the bosons occupy the state A at $k_x=0$.  After the drive is turned on, the bosons still keep the same momentum $k_x=0$, because the kinetic equation (\ref{kinetic_general}) does not include any dissipation and relaxation.  So, their occupation function does not really change, but it now corresponds to the state B with a renormalized energy and a nonzero group velocity.  The question is whether the bosons will stay in the state B forever.  It seems unlikely, and we argue that they will eventually relax to the state D with zero group velocity, when their occupation function $f[\tilde\varepsilon(\bm k)]$ becomes thermalized to the renormalized energy dispersion shown by the dashed parabola.

In principle, dissipation and relaxation can be included in the kinetic equation~(\ref{kinetic_general}).  However, this is often done \textit{ad hoc} by introducing  phenomenological terms proportional to the deviation of the occupation function from a presumed distribution representing thermal equilibrium \cite{Kaplan2020}.  Alarmingly, this approach implicitly introduces an \textit{a priory} assumption toward what state the system is going to relax, which is not entirely obvious for a driven system.  Specifically, a seemingly innocuous assumption that the bosons would relax toward the state of the lowest \textit{unrenormalized} energy $\varepsilon(\bm k)$ would keep them at the state B with $k_x=0$ and a nonzero group velocity.  In contrast, if relaxation is introduced toward the state of the lowest \textit{renormalized} energy $\tilde\varepsilon(\bm k)$, the bosons will move to the state D with zero group velocity and zero current.  It appears that, by introducing a relaxation time $\tau$, Refs.~\cite{SodemannFu2015,Nagaosa2022,Sodemann2022} implicitly make an assumption of the former kind, whereas we argue for the latter.  An experiment with bosons in an optical lattice described in Sec.~\ref{sec:1d_limit} around Fig.~\ref{fig:renorm_dispersion} can settle this dispute.

\end{document}